\documentclass[12pt,preprint]{aastex}
\begin{document}
\def\fas{\hbox{$.\!\!''$}}
\def\msun{$M_{\odot}$}
\def\Rsun{$R_{\odot}$}
\def\msunyr{$M_{\odot}~yr^{-1}$}
\def\phflux{ph~cm$^{-2}$~s$^{-1}$}
\def\enflux{erg~cm$^{-2}$~s$^{-1}$}
\def\phflux2{ph~keV$^{-1}$~cm$^{-2}$~s$^{-1}$}
\def\fluxlam{erg s$^{-1}$ cm$^{-2}$\AA$^{-1}$}
\def\flux{ergs~cm$^{-2}$~s$^{-1}$}
\def\lum{erg~s$^{-1}$}
\def\deg{$^{\rm o}$}

\def\Tin{$T_{\rm in}$}
\def\R*{$R^*$}
\def\Rin{$R^*$}
\def\GHI{$\Gamma_{\rm HI}$}

\def\fdg{\hbox{$.\!\!^\circ$}}
\def\famz{\hbox{$.\!\!'$}}
\def\fas{\hbox{$.\!\!''$}}

\newcommand{\Rs}{{R_{\rm S}}}     
\newcommand{\obs}{{\rm obs}}
\newcommand{\Teff}{T_{\rm eff}}
\newcommand{\Tinf}{T_{\infty}}
\newcommand{\thetac}{\theta_{\rm c}}
\newcommand{\muc}{\mu_{\rm c}}

\shorttitle{X-ray Transient H 1743--322}
\shortauthors{McClintock et al.}

\title{The 2003 Outburst of the X-ray Transient H~1743--322: Comparisons
  with the Black Hole Microquasar XTE J1550--564$^1$}

\author
{Jeffrey E.\ McClintock\altaffilmark{2},
Ronald A.\ Remillard\altaffilmark{3},
Michael P.\ Rupen\altaffilmark{4},
M.\ A.\ P.\ Torres\altaffilmark{2},
\newline
D.\ Steeghs\altaffilmark{2},
Alan M.\ Levine\altaffilmark{3},
Jerome A. Orosz\altaffilmark{5}}

\altaffiltext{1}{This paper includes data gathered with the 6.5 meter
  Magellan Telescopes located at Las Campanas Observatory, Chile.}

\altaffiltext{2}{Harvard-Smithsonian Center for Astrophysics, 60 Garden
Street, Cambridge, MA 02138}

\altaffiltext{3}{MIT Kavli Center for Astrophysics and Space Research,
Massachusetts Institute of Technology, Cambridge, MA 02139}

\altaffiltext{4}{National Radio Astronomy Observatory, New Mexico Array
Operations Center (VLA, VLBA), Socorro, NM 87801}

\altaffiltext{5}{Department of Astronomy, San Diego State University,
5500 Campanile Drive, San Diego, CA 92182}

\email{}
 
\begin{abstract}

The bright X-ray transient H~1743-322 was observed daily by the {\it Rossi X-ray
Timing Explorer (RXTE)} during most of its 8-month outburst in 2003.  We
present a detailed spectral analysis and a supporting timing analysis of
all of these data, and we discuss the behavior and evolution of the
source in terms of the three principal X-ray states defined by Remillard
and McClintock.  These X-ray results are complemented by Very Large
Array (VLA) data obtained at six frequencies that provide quite complete
coverage of the entire outburst cycle at 4.860 GHz and 8.460 GHz.  We
also present photometric data and finding charts for the optical
counterpart in both outburst and quiescence.  We closely compare
H~1743--322 to the well-studied black-hole X-ray transient XTE~J1550--564 and
find the behaviors of these systems to be very similar.  As reported
elsewhere, both H~1743--322 and XTE~J1550--564 are relativistic jet
sources and both exhibit a pair of high-frequency QPO oscillations with
a 3:2 frequency ratio.  The many striking similarities between these two
sources argue strongly that H~1743--322 is a black hole binary, although
presently no dynamical data exist to support this conclusion.

\end{abstract}



\section{Introduction}

Twenty-two X-ray binaries contain a dynamically-confirmed black hole
(Remillard \& McClintock 2006; Orosz et al.\ 2007; Silverman \&
Filippenko 2008).  Spin estimates have been obtained for five of these
black holes (Shafee et al.\ 2006; McClintock et al.\ 2006; Liu et al.\
2008; Miller et al.\ 2008; Reis et al.\ 2008).  These black hole
binaries are important to physics as potential sites for tests of
strong-field general relativity (Remillard \& McClintock 2006; hereafter
RM06).  Seventeen of these systems are transient X-ray sources.  In this
paper, we discuss one such transient system, H~1743--342 (hereafter
H1743).  The mass of its black hole primary has yet to be measured, and
therefore H1743 is not a dynamically-confirmed black hole binary.
However, as we show in this paper, the overall behavior of this
transient X-ray source and its strong similarities to the dynamically
established black-hole transient XTE J1550--564 make a strong case that
H1743 does contain a black hole primary.

The transient H1743 displayed major outbursts in 1977, 2003 and 2008.
In this paper, we present results only for the 2003 outburst cycle.  The
source was initially discovered with the {\it Ariel V} All-Sky Monitor
by Kaluzienski \& Holt (1977), who provided a $\sim1^{\rm o}$ position.
Further definitive X-ray studies were carried out using the instruments
aboard {\it HEAO 1}.  The modulation collimator pinpointed the source
location to one of two equally probable X-ray positions (Doxsey et al.\
1977; Gursky et al.\ 1978).  The high-energy LED detectors observed the
spectrum upward to $\approx 100$~keV (Cooke et al.\ 1984), and the A-2
detectors (1.5--30 keV) were used to measure the medium energy spectrum.
An analysis of these latter data led White and Marshall (1984) to
classify the source as a black hole candidate based on its very soft
spectrum.  During its 1977--1978 outburst, the source reached at least
70\% of the intensity of the Crab (1--10 keV; Doxsey et al.\ 1977;
Kaluzienski and Holt 1977), and the duration of the outburst was at
least seven months (Kaluzienski \& Holt 1977; Cooke et al.\ 1984).

The 2003 outburst was discovered by Revnivtsev et al.\ (2003) on March
21 using {\it INTEGRAL} and rediscovered a few days later by Markwardt
\& Swank (2003), who initiated the first pointed {\it RXTE}
observations of the source on March 28.  Following some early confusion
over the identity of the source (initially dubbed IGR/XTE J17464-3213),
the X-ray transient was correctly identified with the earlier {\it HEAO 1}
source based on its X-ray position (Markwardt \& Swank 2003).  Shortly
thereafter, a radio counterpart was discovered.  The precise position of
the radio source agreed with the {\it HEAO 1} X-ray position to 8$''$
(Rupen et al.\ 2003), which left no doubt that the new
transient and H1743 are one and the same.  Shortly thereafter, a
variable optical counterpart was identified, which is coincident with
the radio counterpart (Steeghs et al.\ 2003; \S3.6).  The source has
been closely monitored at radio wavelengths (e.g., Rupen et al.\ 2003;
\S3.5). 

The discovery of a pair of high-frequency QPOs at 240 Hz and 165 Hz is
an especially important result, which was obtained using {\it RXTE}
timing data (Homan et al.\ 2005; Remillard et al.\ 2006).  Very similar
pairs of high-frequency QPOs with a commensurate frequency ratio of 3:2
are seen for three other dynamical black holes: XTE J1550--564, GRO
J1655--40, and GRS 1915+105 (McClintock \& Remillard 2006).  Thus, the
presence of this distinctive harmonic pair of frequencies for H1743
opens wider the possibility of obtaining fundamental information about
black holes (e.g., Abramowicz \& Kluzniak 2001; Remillard et al.\ 2002a;
Wagoner 1999).  A comparably important discovery is that of the
large-scale relativistic X-ray jets observed for both H1743 (Corbel et
al.\ 2005) and XTE J1550-564 (Corbel et al.\ 2002).  The jets are
observed as plasma blobs, which were presumably ejected during the 2003
outburst of H1743 and the 1998 outburst of XTE J1550-564 (\S4.4).  The
broadband spectrum for both sources is consistent with synchrotron
emission from very high energy particles ($\sim 10$~TeV).

Other work on the 2003 outburst of H1743 includes a report on three
high-energy {\it INTEGRAL} observations (Parmar et al.\ 2003), an
investigation of the low-frequency QPOs (Homan et al.\ 2005), and a
simultaneous {\it Chandra} and {\it RXTE} spectroscopic study (Miller et
al.\ 2006).  Several additional studies of the 2003 outburst cycle of
H1743 have been published.  They are not as comprehensive as the present
work; however, they provide complementary and very valuable information
on such topics as the high energy spectrum, timing behavior, and
accretion-disk winds.  For a thorough and independent spectral/temporal
analysis of the {\it RXTE} data obtained during the last several weeks
of the outburst, with an emphasis on state transitions, see Kalemci et
al.\ (2006).  For a further analysis of state transitions for selected
{\it INTEGRAL}/{\it RXTE} observations, see Joinet et al.\ (2005).  For
additional analyses of {\it INTEGRAL}/{\it RXTE} spectra made at a
number of epochs, see Capitanio et al.\ (2005, 2006) and Lutovinov et
al.\ (2005).

Herein, we present spectral and timing results for 170 pointed X-ray
observations, as well as extensive VLA radio observations.  These
observations span almost eight months.  Also presented are optical and
near-infrared (NIR) data and images for both the outburst and quiescent
states.  This work is a continuation of our program to study the
spectral and timing behavior of Galactic black holes.  Some comparable
studies of transient sources include, e.g., Ebisawa et al.\ (1994) on
Nova Mus 1991; Sobczak et al.\ (1999) on GRO J1655--40; Sobczak et al.\
(2000b) and Remillard et al.\ (2002b) on XTE J1550--564; Park et al.\
(2004) on 4U~1543--47; Belloni et al.\ (2005) on GX~339--4; and Tomsick
et al.\ (2005) on 4U 1630--47.  To these we add the notable study of the
persistent source Cygnus X-1 by Wilms et al.\ (2006).

This paper is organized as follows.  In \S2 we discuss the X-ray, radio,
and optical/NIR observations and data reduction and analysis.  In \S3,
we present our results for all three wave bands, while featuring the
results of our X-ray spectral analysis.  In \S4 we highlight some of
the more interesting behavior exhibited by H1743 while making close
comparisons to the behavior of XTE J1550--564.  Finally, we offer our
conclusions in \S5.


\section{Observations, Reductions and Analysis}

In this section, we discuss the X-ray, radio, and optical/NIR data in
turn.  The X-ray data are described in \S2.1, their reduction and
spectral analysis is detailed in \S2.2 and their timing analysis is
briefly described in \S2.3.  We define and discuss X-ray states in
\S2.4.  Our radio and optical/near-IR observations are described in
\S2.5 and \S2.6, respectively.

\subsection{X-ray Observations}

In Figure~1, we show the X-ray light curve of H1743 for the complete
2003 outburst cycle of the source.  These are data obtained during 170
pointed {\it RXTE} observations using the Proportional Counter Array
(PCA; Jahoda et al.\ 2006).  The intensity and a pair of hardness ratios
are plotted versus ``Day Number'' which is referenced to 2003 March 21
(modified Julian date 52719), the day of discovery of the outburst
(Revnivtsev et al.\ 2003).  We use the Day Number for plots and refer to
it throughout this work.  We also make extensive use of the
``observation number'', which is tabulated along with the Day Number,
the modified Julian date (MJD), and the calendar date for all 170
observations (see Appendix, Tables A1 \& A2).  As indicated in the
figure, observations were performed on an essentially daily basis during
the first 3.5 months when the source was most active, while during the
next 3 months the source was relatively stable and observations were
made only every 3--4 days.  During the several weeks of hard-state
activity that occurred near the end of the outburst cycle, the source
was again monitored on an almost daily basis.  The 170 observations
include all of those obtained in our {\it RXTE} guest observer program
(P80146) plus additional public observations of this source (P80138,
P80144, P80135 and P80137).

\subsection{X-ray Spectral Fitting}

We fitted simultaneously the pulse-height spectra obtained using only
the best-calibrated detector modules of each of the two large-area
instruments on {\it RXTE}, i.e., Proportional Counter Unit 2 (PCU-2) of
the PCA and the four detectors in Cluster A of the HEXTE.  PCU-2 is a
xenon-filled detector with an effective area of $\approx$ 1250 cm$^2$,
an energy range of 3--60 keV, and an energy resolution of $\approx$ 17\%
at 6 keV (Jahoda et al.\ 2006).  The four NaI/CsI detectors of HEXTE-A
provide an effective area of as much as 800~cm$^2$ over the energy range
15--250 keV (Rothschild et al.\ 1998).

The PCU-2 data were taken in the ``standard 2'' format, which consists
of a series of 16-s accumulations of counts that were individually
recorded in 129 pulse-height channels.  The spectra were corrected for
background and a systematic error of 1\% was added in quadrature to the
estimated statistical errors in the count rates.  The latter allows for
uncertainties in the detector response and is generally required for
successful spectral fitting of bright sources.  In like manner, the
standard HEXTE reduction software was used to extract the HEXTE archive
mode data.  The HEXTE modules were alternately pointed every 32~s at
source and background positions, allowing good background subtraction.
We jointly fit each of the 170 pairs of PCA and HEXTE pulse-height
spectra over the energy ranges 2.8--25 keV (PCU-2; PHA channels 4--53)
and 24--200 keV (HEXTE-A; PHA channels 13--56), respectively.  The
energy range adopted for PCU-2 is the current standard choice used to
insure reliable calibration.  For HEXTE-A, using the final spectral
model (see below), we tried several different low-energy limits before
selecting 24 keV, which was the lowest value that allowed nearly all of
the spectra to be well-fitted.  The upper limit of 200 keV allowed us in
every case to capture all of the detected counts without significantly
degrading the signal-to-noise ratio relative to a 100 keV limit.

All the spectra were analyzed using XSPEC version 11.2.0 (Arnaud 1996);
the errors on both the spectral parameters (Table A1) and the fluxes
(Table A2) are consistently given at the 1-$\sigma$ level of confidence.
We experimented with many combinations of spectral components in order
to arrive at a simple model that would give good fits to the maximum
number of spectra.  The basic model that we finally adopted is comprised
of the following three components (McClintock \& Remillard 2006; RM06):
(1) a simple power law ({\sc powerlaw}); (2) a multi-temperature
blackbody accretion disk ({\sc diskbb}; Mitsuda et al.\ 1984; Makishima
et al.\ 1986); and (3) a low-energy cutoff component ({\sc phabs};
Balucinska-Church \& McCammon 1992; Yan et al.\ 1998) with the hydrogen
column fixed at $N_{\rm H}$ = 2.2~$\times 10^{22}$~cm$^{-2}$ (see
below).  In addition, a Gaussian Fe~K emission-line component {(\sc
gaussian)} was added when its inclusion decreased the value of the
reduced $\chi^2$ by more than 0.3\footnote{ Here and elsewhere in this
paper, we use the threshold criterion $\Delta{\chi_{\nu}^{2}} = 0.3$,
which corresponds to a change in the total chi-square of $\approx 25$
for 86--90 degrees of freedom (Table~A1, column 13).  The statistical
importance of including the component in question (here the Gaussian
line) cannot be ascribed a definite statistical significance because our
uncertainties are dominated by the 1\% systematic errors in the detector
response.  If we ignore this problem and use the Akaike Information
Criterion (Akaike 1974) to compare the model with and without the
additional component, we estimate that the statistical significance of
the added component (here the Gaussian line) is $\sim 4-5~\sigma$.}.
The central line energy and line width were allowed to float from
6.4--7.0 keV and 0.3--2.0 keV, respectively.  Interestingly, none of the
fits were improved by the inclusion of a smeared Fe absorption edge
(e.g., Ebisawa et al.\ 1994), which has been widely used in similar
studies (e.g., Sobczak et al.\ 2000b; Park et al.\ 2004), and this
component was not used at all.

In order to achieve satisfactory fits to several dozen of the spectra
that were obtained during the first half of the outburst cycle, it was
necessary to allow for some curvature in the power-law component.  We
examined three approaches, namely (1) a simple power law plus an
exponential cutoff; (2) a broken power law without an exponential
cutoff; and (3) a broken power law with an exponential cutoff.  Approach
(2) gave globally better fit results than approach (1), but compounding
the model further by adding an exponential cutoff did not improve the
fits, and therefore no cutoff was used for the results reported here.
In short, we substituted a broken power-law component ({\sc bknpower})
for the simple power-law component whenever $\chi_{\nu}^{2}$ was thereby
decreased by more than 0.3, which was the case for 48 of the 170
spectra.

The parameters of the broken power law component are a pair of photon
indices ($\Gamma$ and $\Gamma_{\rm HI}$), a break energy ($E_{\rm
break}$), and the normalization (photon flux at 1~keV), all of which
were free to vary with a single constraint, $E_{\rm break} > 12.0$~keV.
The photon index of the power-law component for energies above $E_{\rm
break}$ was allowed to vary independently for the PCU-2 and HEXTE-A
detectors, although the normalization was fitted in common.  As
indicated in Figure~2, the photon indices, $\Gamma$ and \GHI, generally
tracked each other very well over the course of the observations.  The
high-energy photon index \GHI~that we report in \S3 and Table A1 is an
inverse-variance-weighted average of the indices determined
independently by the PCU-2 and HEXTE-A detectors.

In earlier studies of black hole binaries, we and others have found that
the normalizations of the PCA and HEXTE detectors did not agree.  For
example, in our spectral study of GRO J1655--40, the systematic offset
between the two instruments based on observations of the Crab nebula was
$\sim20$\%, and we therefore allowed the HEXTE normalization to float
independent of the PCA normalization with our reported fluxes based
solely on the PCA data (Sobczak et al.\ 1999).  In the present study,
using contemporary response file, we were pleased to find that the
normalizations of the two detectors were in good agreement.  That is, we
fitted all of the spectra using the pairs of detector response files as
generated; i.e., no relative adjustment of the normalizations was made
via a multiplicative constant.

In the case of 16 spectra, which were obtained at the beginning and end
of the outburst cycle, plus several additional spectra near Day 55, the
accretion disk component was excluded from the model because it did not
improve the fit significantly; i.e., $\Delta$$\chi_{\nu}^{2}$ improved by
$<0.3$.  For these spectra, the disk is either too cool to be detected
by the PCA or is not present.

Table A1 lists for each spectrum the fitted values of the spectral
parameters for the three model components: accretion disk (two
parameters in columns 4--5), power law (either two or four parameters in
columns 6--9), and Fe line (three parameters in columns 10--12).  A set
of blanks for a particular component indicates that this component was
excluded from the fit (see above).  The value of the reduced chi-square
and the number of degrees of freedom are given in column 13.  As an
inspection of Table A1 shows, seven spectra (obs.\ no.\ 37, 41, 44 \&
48--51) required the use of all nine parameters in order to achieve an
acceptable fit.  At the other extreme, two spectra (obs.\ nos.\ 165 \&
168) were well fitted using only the two parameters of the simple
power-law component.  Remarkably, with only two exceptions, the 79
spectra corresponding to obs.\ nos.\ 58--136 were well fitted using just
the four parameters that define the disk and simple power-law
components; that is, these pristine spectra required no line/edge
components or spectral break.  Using the models considered here, as well
as alternative models, we were unable to obtain satisfactory fits (i.e.,
$\chi_{\nu}^{2} \lesssim 3$) to three of the 170 spectra (obs.\ nos.\
15, 21 \& 127).

The value of $N_{\rm H}$ was determined in a preliminary analysis of a
number of thermal-state spectra (obs.\ nos.\ 101-134) using the disk and
simple power-law components indicated in Table A1 and by allowing the
low-energy cutoff parameter $N_{\rm H}$ (see above) to vary.  In this
way, we determined an inverse-variance-weighted average value of $N_{\rm
H}$ = 2.2~$\times 10^{22}$~cm$^{-2}$, which we fixed in computing all of
the results reported in Table A1 and elsewhere in this paper.  This
value is very similar to the values reported by others for this source
(Doxsey et al.\ 1977; Parmar et al.\ 2003; Markwardt \& Swank 2003;
Miller et al.\ 2006), and it agrees with the value determined from
optical/NIR observations, $N_{\rm H}$ = 2.1~$\times 10^{22}$~cm$^{-2}$
(\S3.6).  Furthermore, the precise value of $N_{\rm H}$ is unimportant
here because for this column depth the absorption at the 2.8 keV
threshold energy of the PCA is only $\approx 35$\%.

\subsection{X-ray Timing Analysis}

In this paper, our focus is on the analysis and presentation of the
spectral data, and we do not present a full timing analysis of the data
(for examples of such studies, see Remillard et al.\ 2002a on XTE
J1550--564 and Belloni et al.\ 2005 on GX 339--4).)  The timing results
reported here are intended only to supplement our featured spectral
results.  Our search methods for QPOs have been described previously in
Remillard et al.\ (2002a, 2002b), and we refer the reader to these
papers for background on how the power density spectrum (PDS) is
computed, normalized, corrected for dead time, binned and averaged and
how our error estimates are determined.  The method of detecting QPOs is
likewise described in these references.  Briefly, we use a sliding
frequency window, which typically spans 0.2 $\nu_{\rm trial}$ to 5.0
$\nu_{\rm trial}$.  Within this window, the power continuum ($P_{\rm
cont}$) is modeled with a second order polynomial in log $P_{\nu}$
vs. log $\nu$, since the local power continuum usually resembles a
power-law function with a slight amount of curvature. The QPO profiles
are presumed to be Lorentzian functions, and they are generally
distinguished from broad peaks in the power continuum by a coherence
parameter, $Q = \nu / FWHM \ga 2$. We use $\chi^2$ minimization to
obtain the best fit for the QPO profile and the local power continuum.
When QPO detections are significant, we use the best-fit values for the
Lorentzian peak and FWHM to compute the total integrated power ($P$) in
the PDS feature; the rms amplitude of the QPO is then $a = P^{0.5}$.

All of the low-frequency QPO results reported in this paper pertain to
power density spectra computed for the full bandwidth of the PCA
instrument (effectively 2--40 keV) in the frequency range 4 mHz to 4
kHz.  Sometimes low-frequency QPOs can appear with harmonic features,
e.g., with frequencies in the ratio 1:2 and sometimes 0.5:1:2 (i.e.,
with a very weak subharmonic; see Remillard et al.\ 2002a).  In such
cases we force the harmonic ratios to be exact and vary only the central
frequency, while allowing the widths of each feature to vary
independently.

Because the timing analysis is secondary and supplementary to the
spectral analysis, we average the QPO properties over the same 170 time
intervals used for the spectral analyses; i.e., we do not examine in
detail those few cases for which the timing properties of the source
varied during the course of a single observation.  Low-frequency QPOs
were detected in a total of 87 of these data sets, and the QPO
parameters (central frequency $\nu$, amplitude $a$, and quality factor
$Q$) are reported in Table~A2.  Also given in Table~A2, in this case for
all 170 observations, is the total rms power ($r$) integrated over
0.1--10 Hz (2--30 keV).

\subsection{X-ray States of Black Hole Binaries}

Extensive observations with {\it RXTE} have shown that each of the
commonly identified states of black hole binaries may characterize the
spectral and timing characteristics of at least some of the systems over
a range of two or more decades in X-ray luminosity.  Therefore,
luminosity is a poor criterion to use in defining the outburst states.
We thus use the state names introduced by McClintock and Remillard
(2006) for the three principal states: namely, thermal dominant (TD;
cf., high soft), hard (H; cf., low hard), and steep power law (SPL; cf.,
very high).  Note that herein and in RM06, we generally refer to the
thermal dominant state as simply the thermal state.  Because these
revised state names are relatively unfamiliar, we list and define them
in Table~1 in terms of the specific characteristics of the energy and
power density spectra.  The state definitions in the table are identical
to those given by RM06, and as indicated in footnotes {\it e} and {\it
f}, they differ slightly from the original definition given by
McClintock \& Remillard (2006).  As noted in RM06, these minor revisions
resulted from gaining additional experience in analyzing long intervals
of data for many black hole binaries.

The three principal outburst states defined in Table~1 represent an
attempt to empirically define limited ranges of spectral and timing
characteristics that often endure for prolonged periods of time and that
appear to have distinct physical origins.  There are gaps in the
parameter ranges used to define these states, and this gives rise to
intermediate or hybrid states that lie between these three principal
states.  In these cases, when possible we specify the admixture of the
pair of principal states that describes the intermediate state in
question (e.g., H:SPL).

\small
\begin{center}
\begin{tabular}{ll}

\multicolumn{2}{c}{Table 1. Outburst States of Black Holes: Nomenclature
and Definitions} \\ \\



\hline
\hline
{\bf New State Name}        &Definition of X-ray State$^a$ \\
(Old State Name)            & \\

\hline

{\bf Thermal}               &Disk fraction$^b$ $>$ 75\% \\
(High/Soft)                 &QPOs absent or very weak: $a^c$ $<$ 0.005 \\
                            &Weak power continuum: $r^d$ $<$ 0.075$^e$ \\
\hline
{\bf Hard}                  &Disk fraction$^b$ $<$ 20\% (i.e., Power-law fraction$^b$ $>$ 80\% \\
(Low/Hard)                  &$1.4^f < \Gamma < 2.1$ \\
                            &Strong power continuum: $r^d$ $>$ 0.1 \\
\hline
{\bf Steep Power Law (SPL)} &Presence of power-law component with $\Gamma > 2.4$ \\
(Very high)                 &Power continuum: $r^d$ $<$ 0.15 \\
                            &Either disk fraction$^b$ $<80\%$ and 0.1-30 Hz QPOs present with $a^c$ $>$  0.01 \\
                            &or disk fraction$^b$ $<$ 50\% with no QPOs present. \\
\hline
\multicolumn{2}{l}{$^a$2--20 keV band.} \\
\multicolumn{2}{l}{$^b$Fraction of the total 2--20 keV unabsorbed flux.} \\
\multicolumn{2}{l}{$^c$QPO amplitude (rms).} \\
\multicolumn{2}{l}{$^d$Total rms power integrated over 0.1--10 Hz.} \\
\multicolumn{2}{l}{$^e$Formerly $r < 0.06$ in McClintock \& Remillard 2006.} \\
\multicolumn{2}{l}{$^f$Formerly $1.5 < \Gamma$ in McClintock \& Remillard 2006.} \\
\end{tabular}
\end{center}
\normalsize

\subsection{VLA Radio Observations}

We observed H1743 a total of about 50 times throughout the course of the
outburst at 8.460 GHz and 4.860 GHz and about half that many times at
1.425 GHz.  The radio data were reduced using the Astronomical Image
Processing System (AIPS).  During the first month of the outburst, about
a dozen observations were made at each of three higher frequencies:
14.94 GHz, 22.46 GHz and 43.34 GHz.  All of the radio data and the
corresponding VLA-antenna configurations are summarized in Tables
A3--A5.  The source was observed most frequently at 8.460 GHz and was
reliably detected through Day 114; thereafter, our observations yielded
only upper limits on the source flux density ($S_{\nu} \lesssim 0.3$
mJy).  The peak flux density observed at 8.640 GHz was 68 mJy on Day
18.5.

\subsection{Optical and Near-infrared Observations}

Optical images were obtained using the MagIC CCD camera mounted on the
Magellan-Clay telescope at Las Campanas Observatory. This imager
provides a $2\famz35$ field of view sampled at
$0\fas069~$pixel$^{-1}$. On 2003 April 5.4 UT (Day 15.4 of the
outburst), 300 s Johnson $R$- and $I$-band images were obtained under
seeing conditions of $0\fas48$. We also secured 5x300~s $i'$-band
exposures of the quiescent state on 2006 June 23.1 UT under photometric
conditions and with a seeing of $0\fas43$. The frames were de-biased and
flatfielded and an astrometric frame in the ICRS J2000.0 system was
calculated using a large number of 2MASS (Skrutskie et al.\ 2006) and
UCAC2 reference stars.

Dithered $K_{\rm s}$-band images, also of the quiescent state, were
obtained during a total of 3 hours using the PANIC camera on the
Magellan-Baade telescope on the nights of 2006 May 7--8 UT.  PANIC
(Persson's Auxiliary Nasmyth Infrared Camera: Martini et al.\ 2004) is a
near-infrared camera with a pixel scale of $0\fas125$pixel$^{-1}$ that
projects a $2'$ field of view onto a Rockwell 1024x1024 HgCdTe detector.
Observations consisted of a 5-point dither pattern with a 25 s exposure
that was repeated three times at each dither position, yielding a total
of 6.25 min on source. The field of H1743 is crowded, and therefore each
pair of 6-min dither sequences on the object was followed by a dithered
observation of a relatively blank region of sky. These sky-blank frames
were used to subtract the sky background from the target frames.  We
also observed the standard star S279-F (Persson et al.\ 1998), which we
used for the absolute flux calibration.  The reductions were done using
PANIC/IRAF software. Seeing was typically $0\fas5$. In addition to these
quiescent images, $K_{\rm s}$-band images of H1743 were obtained with
the PANIC camera near the end of the 2003 outburst on Day 175.1 (2003
September 12.1 UT). The same dither pattern was used with exposures of
10 s, giving 2.5 min on source in seeing of $0\fas5$.


\section{Results}

The PCA light curve shown in Figure~1 provides a global view of the
outburst. It reveals that during the first three months H1743 flared
violently and exhibited a generally hard spectrum.  Thereafter, the
flaring subsided, and for several months the intensity of the source
decayed smoothly and the spectrum was quite soft.  During the final few
weeks of decay, the source spectrum hardened.

We now examine the outburst in greater detail. The spectral parameters
given in Table~A1 and the fluxes and timing data listed in Table A2 were
used in concert with the state definitions given in Table~1 to determine
the state of the source during each of the 170 observations.  Our state
classifications are listed in the far-right column in Table~A2, and they
are displayed as color-coded symbols in Figure~3$a$, which is a plot of
the 2--20 keV model flux computed using the spectral parameters given in
Table~A1.  The ASM light curve is shown in Figure 3$b$; note the clear
delineation it provides of the initial rise of the outburst.

Figure~3$a$ provides a summary view of how the state of the source
evolved during the course of the outburst.  The source started and ended
the outburst in the hard state (blue squares).  Furthermore, on one
other occasion during the outburst, namely Day 58, the source returned
briefly to the hard state, and around this time a thermal component was
not detected during five observations (Table~A1).  During the most
intense period of flaring, the source was generally in the SPL state
(green triangles), the period when high-frequency QPOs were detected
(Homan et al.\ 2005; Remillard et al.\ 2006).  Interestingly, in the
middle of the flaring period (Days 39--43), the source entered a
thermal-state lull for five days.  Commencing on Day 88, the source
again transitioned into the thermal state (red crosses).  Thereafter it
underwent a gradual and steady 4-month decay before subsiding into the
hard state via the H:SPL intermediate state.  The occurrence of
intermediate states (magenta circles), which were common during the
flaring phase and again near the end of the outburst, are all hybrids of
the hard and SPL states (H:SPL) with three exceptions: On Days 44, 64
and 95 the source was found in a state intermediate between the thermal
and SPL states (TD:SPL; see Table~A2).  We note that the first
observation on Day~7.8 with its hot ($kT_{\rm in}=1.78$~keV), tiny
($R^*_{\rm in}=1.2$~km) thermal component cannot be accommodated within
our state classification scheme, as indicated in the last column of
Table~A2.

This section is organized as follows.  In \S3.1 we present and discuss a
pair of figures that illustrate the principal spectral and timing data
contained in Tables A1 and A2.  In \S3.2 we show representative energy
spectra for several of the observations.  The thermal state is the
subject of \S3.3.  In \S3.4 we examine correlative relationships between
the spectral and temporal data and also discuss an impulsive power-law
flare observed on Day 47.  In \S3.5 we present the radio data and
discuss their connections with the X-ray data, and finally, in \S3.6 we
present the optical/NIR results for both the outburst and quiescent
states.

\subsection{Spectral and Temporal Evolution of the Outburst}

Figure~4, which displays most of the data contained in Tables A1 \& A2,
offers a detailed look at the full spectral evolution of the source.
Panel $a$ indicates the state of the source (\S2.4).  The spectral
parameters are presented in panels $b-d$: the disk blackbody temperature
($b$), the inner disk radius ($c$), and the low- and high-energy photon
indices ($d$).  Flux data are shown in the five panels in the lower
portion of the figure: the Fe K emission line flux ($e$); the disk,
power-law, and total fluxes in the 2--20 keV band ($f-h$); and the flux
in the 20--100 keV band ($i$).  The complex behavior of the source
during the first few months and its stable nature thereafter is as
apparent in this record as it is in the light curves shown in Figure~3.

Figure~5 presents a companion view of the outburst cycle that reveals
its temporal evolution.  Shown are ($a$) the X-ray state, ($b$) the rms
power, $r$, integrated over 0.1--10 Hz in the 2--30 keV band, ($c$) the
central frequency of the dominant QPO, ($d$) its amplitude, and ($e$)
its quality factor $Q = \nu/\Delta\nu$.  For comparison, the Fe K line
and disk fluxes are repeated from Figure~3 in ($f$) and ($g$),
respectively.  In this record, low-frequency QPOs and a strong power
continuum are much in evidence throughout most of the first few months.
They disappear for the next few months, while the source remains fixed
in its thermal state, and they then reappear near the end of the
outburst as the spectrum hardens.

The outburst cycle can be divided into three main intervals.  The first
of these is the initial 87 days, which is marked by strong flaring
activity and a generally dominant power-law component.  During the
second interval, the source remained locked in the thermal state for 125
days (through Day 211) as its intensity gradually and monotonically
decayed.  During the last interval of 22 days, the spectrum hardened as
the source made its return to the quiescent state.  We now discuss these
time intervals in turn in relation to Figures~4~\&~5.

{\it Flaring Period -- the first 87 days:} As indicated in Figure~4,
during Days 8--23, the source transitioned trom the hard state through
an intermediate H:SPL state to the SPL state.  During this period the
disk flux is very low, the power-law flux rises steadily, the Fe line
flux is solidly detected in four observations, there is a large
difference between the two spectral indices, and QPOs are present
throughout (Fig.\ 5).  Days 24--38 mark a period of intense and
sustained flaring (see also Fig.\ 3b).  During this period the source is
predominantly in the SPL state, the disk flux and temperature is high,
the power-law indices become less discordant, and the Fe line is
undetected; the QPO frequency is high and the amplitude is generally
low.  Throughout Days 39--43, the source remains in a thermal-state
lull.  Then, during Days 44--64, the disk flux plummets and remains
low. The source initially alternates between the SPL and the hard:SPL
states, later establishes itself in the SPL state and briefly visits the
hard state on Day 57.7.  An intense, impulsive flare is seen on Day 46.9
during which the power-law flux triples (Figs.\ $4g$ \& $i$; \S3.4.3).
The power-law flux remains high, the two photon indices diverge
strongly, the Fe line is in evidence, and the QPO amplitude is high.
Days 65--87 mark the second flaring phase, which is dominated by the SPL
state.  Three additional power-law flares are seen on Days 67.3, 75.6
and 79.5 (Figs.\ $4g$ \& $i$), the disk temperature declines, and the
inner disk radius increases.  QPOs of high frequency and low amplitude
are present, and the power continuum plummets to a low value (Fig.\ 5).

{\it Thermal period -- Days 88--211:} With the exception of Day 94.8,
the source remains in its thermal state, QPOs are absent, and the
power continuum is at its rock bottom level.  During the entire
interval, the disk temperature falls gradually and monotonically, and
the inner disk radius is fairly constant, increasing smoothly by only
$\approx 25$\%.  The Fe line flux is undetectable for most of this
4-month period, making an appearance only during the last few weeks.
The 2--20 keV power-law flux undergoes an initial rapid decay and then
remains consistently low (Fig.\ $4g$).  For an extended interval prior
to Day 170, the 20--100 keV power-law flux is extremely low,
contributing only an average of 0.3\% of the total flux (Table A2; Fig.\
$4i$).  Around Day 170, however, the intensity of the hard flux
increases suddenly by an order of magnitude and remains at this level of
intensity on average through to the end of the thermal period.  Prior to
Day 102, the power-law index is relatively hard: $\Gamma \approx 2.3$.
However, on Day 102, the index steepens dramatically from 2.6 to 3.5 in
a single day (Fig.\ 5; Table A1).  This steep power-law slope is
maintained for about two weeks, and then a period of erratic variability
sets in for almost ten weeks, during which the index varies from
$\approx 2.3-3.8$.  In the final few weeks, the index again achieves a
stable value, although it has now again become quite hard: $\Gamma
\approx 2.2$.

{\it Transition to the Hard State -- Days 212--233:} During the first
week of this period, the source is in the H:SPL intermediate state.
During this time, QPOs reappear and the power continuum climbs steadily.
Shortly thereafter, the source arrives at the hard state and the power
continuum reaches a local maximum of $r=0.18$ on Day 224 (Fig.\ $5b$).
During this same period, the QPO frequency declines steadily from an
initial maximum of 7.9 Hz to a minimum of 2.5 Hz, and the amplitude
increases to a maximum of 7.6\% (Table A2).  The power-law flux, which
reemerged at the end of the thermal period, strengthens further,
reaching a maximum around Day 220.  During the final 10 days of this
period, the source settles into the hard state and its intensity
declines steadily, falling to a level of $\approx 7$ mCrab on the final
day (2--20 keV; Table A2).  The QPOs shift to lower frequencies,
becoming undetectable by Day 226.  During the final two weeks, the Fe
line flux declines and, during the final week, the disk component
becomes undetectable.

\subsection{Representative Energy Spectra}

Eight energy spectra that illustrate the range of spectra observed for
H1743 are shown in Figure~6.  Initially, on Day 18.6, a strong power-law
component dominates the spectrum and a significant and narrow Fe line
component is present.  On Day 41.8, the thermal component is ascendent
for $E < 10$~keV, although a hard and intense power-law tail is also
present.  The SPL spectrum for Day 46.9 shows the dominant power-law
component that corresponds to the intense flare observed on that day
(see \S3.4.3).  On Day 57.7, the disk temperature falls below the
threshold for detection, and a hard-state spectrum extending to $\approx
170$ keV is observed.  The SPL spectrum observed on Day 71.1 corresponds
to the second epoch of flaring activity.  The thermal spectrum for Day
110.3 was obtained three weeks after the onset of the decay of the disk
emission; note the faint and steep power-law component.  The spectrum
for Day 193.5 shows another example of thermal-state behavior that was
observed late in the decay of the disk; note the similar thermal-state
spectrum for Day 41.8 and the dissimilar one for Day 110.3; note also
the presence of a broad Fe line component.  During the last 10 days of
the outburst (Day Numbers $> 223$), the spectrum is in the hard state.
However, the final observation shown here for Day 233.3 is classified as
H:SPL because the power-law index was slightly too soft ($\Gamma = 2.14
\pm 0.02$) to permit a hard-state classification (see footnote $g$ in
Table~A3).

\subsection{Thermal State: Comparison with XTE~J1550--564}

For many years, the thermal state of black-hole X-ray binaries has been
modeled successfully using a simple multi-temperature accretion disk
model (e.g., Shakura \& Sunyaev 1973; Mitsuda et al.\ 1984; Makishima et
al.\ 1986).  The significant successes of this approach derive from the
relative simplicity of this state for which the power-law component is
weak, QPOs are absent or very weak, and the power continuum is faint
(\S2.4).  An important result that illustrates the simplicity of this
state is summarized in the review by Tanaka \& Lewin (1995; see
references therein).  They show examples of the steady decay (by factors
of 10--100) of the thermal flux of three sources during which $R_{\rm
in}$ remains quite constant (see their Fig.\ 3.14).  More recently, this
evidence for a constant inner radius in the thermal state has been
presented for a number of sources via plots showing that the bolometric
luminosity of the thermal component is approximately proportional to
$T_{\rm in}^4$ (Kubota et al.\ 2001; Kubota \& Makishima 2004;
Gierli\'nski \& Done 2004; Abe et al.\ 2005).

In Figure 7, we show such plots of the bolometric luminosity of the
thermal component versus the color temperature $T_{\rm in}$ for the
thermal-state data for H1743 and XTE~J1550--564; the latter plot is
based on the tabulated results in Sobczak et al.\ (2000b).  These
luminosities are averages over $4\pi$ steradians assuming that the
observed thin-disk flux is $\propto$~cos$i$ (i.e., ignoring limb
darkening), and they are based on a distance of $D=5.5$ kpc and an
inclination of $i=70$\deg~for XTE J1550--564 (Orosz et al.\ 2002) and
assumed values of $D=10$ kpc and $i=70$\deg~for H1743.  (Luminosities
quoted elsewhere in this paper are based solely on the observed flux.)
Very similar plots for XTE J1550--564 are shown in Gierli\'nski \& Done
(2004) and Kubota \& Makishima (2004).  For both sources, Figure~7 shows
that the bolometric disk luminosity rises with temperature more slowly
than $T_{\rm in}^4$, or, equivalently, the color radius $R_{\rm in}$
decreases somewhat with increasing temperature.  For example, during the
4-month decay in the thermal state (Days 88--211) the radius increases
by $\approx 25$\% (Fig.\ 4$c$; Table A1).

\subsection{SPL, Hard and Intermediate States}

We now present results for phenomena that are absent or only weakly
present in the thermal state, namely QPOs, the 0.1--10 Hz power
continuum, and the power-law component.  We discuss the following three
topics in turn: (1) The timing data contained in Table A2; (2) for both
H1743 and XTE J1550--564, we show four selected correlation plots that
relate the various spectral and timing parameters; and (3) we compare
the strongest of the impulsive power-law flares observed for H1743 to
the unprecedented 7-Crab flare that was observed during the 1998--1999
outburst of XTE J1550--564.

\subsubsection{Low-Frequency Quasi-Periodic Oscillations}

QPOs were detected for 87 of the 170 observations (Figure~5; Table A2).
In Table A2, in addition to presenting the central frequency, amplitude
and quality factor $Q$ of the dominant QPO, we provide extensive
footnotes describing the harmonic content, if any, of each dominant QPO,
and we also note the presence of any additional QPOs.  In this paper, we
do not show power spectra, but illustrative examples of spectra for our
comparison source XTE J1550-564 can be found in Figures 1--3 in
Remillard et al.\ (2002b).  There one will see examples of power spectra
with a single dominant QPO and with harmonic components present at 0.5
and twice the central frequency, as described for H1743 in (e.g.)
footnotes $h$ \& $j$ in Table A2.  One will also find in Remillard et
al.\ an extensive discussion of phase lags, QPO types and other
properties of low-frequency QPOs, which is beyond the scope of this
work.

\subsubsection{Spectral/Temporal Correlations}

For the 87 observations for which QPOs were detected, in Figures~8--11
we present selected correlation plots relating the QPO frequency, disk
flux, 3--15 keV photon index, QPO amplitude and continuum power density
for both H1743 and XTE J1550--564.  For the latter source, the spectral
and QPO data were obtained from Sobczak et al.\ (2000b) and Remillard et
al.\ (2002b), respectively, except for the rms continuum power data,
which we computed for this work.  A wider assortment of such correlation
plots have been presented previously for the black hole X-ray transients
XTE J1550--564 and GRO J1655--40 by Sobczak et al.\ (2000a) and
Remillard et al.\ (2002b), and we refer the reader to these works for
further detail.  Furthermore, for the two sources in question, several
additional correlation plots, including correlations that relate count
rates to hardness ratios can be found in RM06 (see their Fig.\ 7 for
H1743 and their Fig.\ 6 for XTE J1550--564).

Figure~8 shows the QPO frequency plotted versus the disk flux for H1743
and XTE~J1550--564.  The striking feature of these plots are the very
similar, tight and approximately linear correlations that exist at
relatively low values of the disk luminosity as the source transitions
between the hard and SPL states via the intermediate H:SPL state.  This
behavior is most apparent for XTE J1550-564, where a very tight
correlation exists for $\nu \lesssim 7$~Hz and disk flux $ \lesssim 1.0
\times 10^{-8}$~\flux.  Note that the frequency range for this
correlation is about the same in the case of H1743.  For both sources,
one significant feature of these strongly correlated data is that the
power-law flux is typically about an order of magnitude greater than the
disk flux.

Figure~9 shows the QPO frequency plotted versus the photon index
$\Gamma$.  Figure~10 shows that the continuum power density is strongly
anti-correlated with QPO frequency.  Figure~11 shows a weaker
anti-correlation between QPO amplitude and QPO frequency.  The amplitudes
are quite high at low frequencies and quite comparable for the two
sources.

In all of the correlation plots shown, there are a number of quite
distinct clusters of points.  A detailed timing analysis would show that
these groupings are associated with different types of low-frequency
QPOs that can be distinguished by their coherence and their phase lag
spectra.  For the definitions of the various QPO types and a detailed
timing analysis of the data for XTE J1550-564 that is discussed herein,
see Remillard et al.\ (2002b).

\subsubsection{Impulsive Power-Law Flares}

Brief flares in the power-law flux were observed in single PCA
observations on Days 32.3, 46.9, 67.3, 75.6 and 79.5 (Fig.\ $4g$ \&
$4i$).  All of these flare events occurred in the SPL state.  During the
strongest of these events, which occurred on Day 46.9, the 2--20 keV
flux increased by a factor of $\approx 2.5$, and the 20--100 keV flux
increased by a factor of $\approx 3$ (Fig.\ $4g$ \& $4i$; Table A2).
The increases in power-law flux are accompanied by corresponding
apparent decreases in the inner disk radius (Fig.\ $4c$) and increases
in the disk temperature (Fig.\ $4b$).

It is of interest to compare the giant flare observed for XTE J1550--564
on 1998 September 19--20 (Sobczak et al.\ 2000b) with the most intense
flare observed for H1743, which occurred on Day 46.9.  For both sources,
the flares occurred in the SPL state (RM06).  Superficially, these
flares appear quite similar.  For example, for both flares the power-law
flux increased by the same factor of about 2.5 (Table A2; Sobczak et
al.\ 2000b), and we note that both flares were observed only during a
single, daily PCA observation.

As shown in Figure~12, in the case of XTE J1550--564 the impact of the
giant flare on the spectrum of the thermal component was far greater
than that observed for H1743.  For example, in the former system
the apparent inner radius of the disk decreased by a factor of 16 during
the flare and the apparent disk temperature quadrupled, reaching $kT =
3.3$~ keV.  The variations in the case of H1743 are much smaller:
The fitted values of the inner disk radius decreased by a factor of
$\approx 2$, and the disk temperature rose by only $\approx 40$\%.  The
behavior of the disk and Fe-line fluxes, which are quite different for
the two sources, are also compared in Figure~12.

We placed limits on the durations of these flares by examining the ASM
count rate data within 10 days of the flare events.  The ASM detected
the giant flare of XTE J1550--564 during eight dwells, which bracket the
time of the PCA observation; these data establish that the flare's
duration was at least 0.87 days, but not more than 2.0 days.  The H1743
flare event was not captured by the ASM, but it was recorded throughout
the 3.4 ks PCA observation.  We thus conclude that its duration was at
least 0.04 days, but not more than 1.55 days.  

\subsection{VLA Radio Results: Comparisons with X-ray Data}

Measurements of the radio flux densities at 1.425, 4,860 and 8.460 GHz,
which span the entire outburst cycle, are shown in Figure~13$b$, and the
state of the X-ray source is indicated above in panel $a$.  Strong radio
flares approaching a flux density of 100 mJy were observed during the
first half of the outburst, whereas only upper limits were obtained
during the second half.  Observations at 14.940 GHz and two higher
frequencies were made during the first 37 days of the outburst.  These
data, which are shown in Figure~13$c$, also exhibit strong flaring
behavior.  The radio spectral index $\alpha$ is plotted in panel $d$.
For comparison, the 2--20 keV power-law component of the X-ray flux is
shown in Figure~13$e$.  The radio flux density fades to $\sim 1$ mJy on
Day 80, while the X-ray source is still in the SPL state.  The radio
intensity further decays to the detection limit (see Table A3) by Day
117, which is $\approx 4$~weeks after the source left the SPL state and
entered the thermal state.

A closer comparison between the radio flux density (red triangles) and
certain X-ray quantities (black filled circles) is displayed in
Figure~14, which covers only the first 100 days of the outburst when the
radio/X-ray source was most active.  As usual, the state of the X-ray
source is indicated in panel $a$.  The X-ray quantities shown are ($b$)
the 2--20 keV power-law flux, ($c$) the difference between the
high-energy and low-energy photon indices ($\Gamma_{HI} - \Gamma$), and
($d$) the Fe K line flux.  These X-ray quantities were selected because
they appear to correlate best with the behavior of the contemporaneous
radio data, which are shown superimposed and arbitrarily scaled in each
of the three panels $b-c$.  For clarity, only the 8.460 GHz radio data,
which was the most frequently sampled band, are overplotted on the X-ray
data.  All of the radio data plotted in Figures~13~\&~14 are given in
Tables A3--A5.  Further discussion of the VLA results is given in \S4.5.

\subsection{The Optical and Near-infrared Counterpart}

In Figure~15, we compare optical observations of the field in quiescence
($a$) and in outburst ($b$).  The latter observations were made early in
the 2003 outburst cycle on Day 15.4 and show a counterpart consistent
with the VLA radio position at a brightness level of $I = 19.3$ and $R =
21.9$ (Steeghs et al.\ 2003).  These magnitudes plus the $K$-band
magnitude of $K_{\rm s} = 13.9 \pm 0.2$ obtained on the same night (Baba
\& Nagata 2003) suggest a reddening that is consistent with the Schlegel
et al.\ (1998) estimate of $E_{\rm B-V} = 3.5$ mag, which implies
$N_{\rm H} = 2.1 \times 10^{22}$ cm$^{-2}$ for $A_{\rm V} = 11.6$ mag
(Schlegel et al.\ 1998; Predehl \& Schmitt 1995).  This is essentially
the same value we obtained by analyzing the {\it RXTE} data: $N_{\rm H}
= 2.2 \times 10^{22}$ cm$^{-2}$ (\S2.2).  Using the extinction law of
Schlegel et al., this reddening estimate suggests dereddened magnitudes
of $I_{0} = 12.5$ and $R_{0} = 12.6$ on Day 15.4 (April 5.4 UT). The
position of the source in the ICRS frame is
$\alpha$(J2000)=17$^h$46$^m$15$^s$.6 and $\delta$(J2000)=--32$^{\rm
o}$14$'$00$''$.9, whose accuracy is limited by the 0\fas25 rms of the
astrometric solution.  We were unable to detect the source in quiescence
during deep $i'$-band exposures on 2006 June 23 (Fig.\ $15a$), implying
a quiescent magnitude limit of $i'>24$. This in turn implies an outburst
amplitude in the optical of over 4.5 magnitudes.  At the same time, our
dereddened magnitudes above, our assumed distance estimate of 10 kpc
(\S3.3), and an assumed value of $(V-R)_{0} = -0.1$ imply a quite
luminous counterpart with $M_{\rm V} = -2.5$ (e.g., see van Paradijs \&
McClintock 1994).

Figure~$16a$ shows the best $K_{\rm S}$ image (0\fas43 seeing) that we
obtained on 2006 May 7.4 UT.  The location of the quiescent counterpart
is consistent with the optical and radio positions (Rupen et al.\ 2003).
PSF photometry (Stetson 1987) was performed to derive a quiescent
magnitude of $K_{\rm s} = 17.1 \pm 0.1$.  The comparison $K_{\rm
S}$-band image in Figure~$16b$ shows the counterpart as it appeared near
the end of the outburst on Day 175.1 (2005 September 12.1 UT).


\section{Discussion}

\subsection{The Evolution of H1743 and other X-ray Transients}

The spectral and temporal evolution of H1743 (Figs.\ 1, 3--5) is
markedly similar to that observed for XTE J1550--564 in 1998--1999
(Sobczak et al.\ 2000b; RM06) and for GRO J1655--40 in 1996--1997
(Sobczak et al.\ 1999; Remillard et al.\ 2002b; RM06).  The outburst
cycles of all three sources show a double-peaked profile.  During the
first maximum, each source primarily exhibited strong flaring behavior
and intense nonthermal emission, whereas during the second maximum the
source spectrum was generally soft and thermal (RM06).  H1743 deviated
somewhat from this pattern because the second maximum begins in a
strongly nonthermal state.  Nevertheless, the outburst cycles of all
three sources followed the same general sequence of events: (1) a rapid
rise to maximum, (2) several months of strong flaring activity, (3)
several months of slow decay in the thermal state, and (4) a final stage
of rapid decay into quiescence.

An even more complex evolution was observed for the recent outburst of
the recurrent X-ray transient 4U1630--47.  This source flared strongly
throughout its long outburst and did not display a lengthy period of
decay in the thermal state (Tomsick et al.\ 2005).  As is well known, a
significant number of other X-ray transients display simple light curves
that rise rapidly, decay exponentially, and are relatively devoid of
flaring behavior.  These include such sources as A0620--00, Nova Mus
1991, GS2000+25, and 4U~1543--47 (Tanaka \& Lewin 1995; Park et al.\
2004).  It has been suggested that the relative complexity of the light
curves of X-ray transients may in part be caused by inclination effects
(Narayan \& McClintock 2005).

For a detailed discussion of physical models of the three active
emission states that considers emission mechanisms, geometry, results
from multiwavelength studies, etc., see \S7 in RM06.

\subsection{The Intermediate State: Fe K line and Spectral/Temporal 
Correlations}

Figure 5 shows that during the first half of the outburst the H:SPL
intermediate state is generally characterized by maximal values of the
QPO amplitude ($\approx 12$\%) and Fe line intensity.  The widely
accepted view is that the Fe line is fluoresced in the cold accretion
disk by the hard power-law flux.  It is then interesting to note that
although the intensity of the power-law component is generally higher in
the SPL state, particularly during the power-law flares (\S3.4.2), than
in the H:SPL state, nevertheless the Fe line is not detected in the SPL
state (see Table A1 \& A2).  Perhaps the higher efficiency of the
intermediate state for producing the Fe line is due to a favorable
geometry or is related in some way to the large amplitude of the QPOs.

At frequencies in the PDS below $\approx 7$ Hz, H1743 and XTE J1550--564
exhibit very similar and nearly linear correlations between the QPO
frequency and disk flux (Fig.\ 8) that extend over more than a decade in
both frequency and flux.  Similar quasi-linear relations between these
two quantities have been observed for other sources, such as for the
1--15 Hz QPOs in GRS~1915+105 (Markwardt et al.\ 1999; Muno et
al.\ 1999) and the 20--30 Hz QPOs in XTE~J1748-288 (Revnivtsev et al.\
2000).  In addition, correlations between the photon index and QPO
frequency (Fig.\ 9) have been extensively studied by Vignarca et al.\
(2003), Titarchuk \& Fiorito (2004) and Titarchuk \& Shaposhnikov
(2005).  The ubiquity of such correlations for black hole binaries and
the striking similarities in the behaviors of H1743 and XTE J1550--564
(Figs.\ 8--11) motivate us to examine more closely these strongly
correlated data.

As stressed by Sobczak et al.\ (2000a) and others, the tight correlation
between QPO frequency and disk flux suggests that the accretion disk
regulates the frequency of the QPO oscillations and that the QPO
frequency is slaved to the rate of mass accretion through the inner
disk.  However, as Sobczak et al.\ and others also stress, the QPO
phenomenon is closely related to the presence of the power-law
component, which must reach a threshold of $\sim 20$\% of the total
(2--20 keV) flux in order to trigger the variable low-frequency QPOs.
For H1743 we find a similar but higher threshold of $\sim 40$\% (Table
A2).  However, the power-law component is more than simply a trigger: It
must also participate directly in the oscillations because when the QPO
amplitude is large ($\sim 10-15$\%), the rms amplitude often exceeds the
fraction of the total flux supplied by the disk alone (Table~A2).  Also,
the energy spectrum of the QPOs is generally hard, indicating that the
bulk of the flux is derived from the power-law component (Muno et
al.\ 1999; Sobczak et al.\ 2000b).  For the strongly correlated data, it
is important to note that the power-law component is totally dominant:
The power-law flux is $\gtrsim 80$\% of the total flux when the QPO
frequency and disk flux are strongly correlated.  This is true for both
H1743 and XTE~J1550--564. Thus, although it is the disk flux that
figures in the tight correlations shown in Figure~8, nevertheless, the
bulk of the total flux is supplied by the power-law component.

\subsection{The Thermal State and the Inner Disk Radius}

In Figure~7, the bolometric disk luminosity is seen to deviate from the
simple $T_{\rm in}^4$ dependence for both XTE J1550--564 and H1743.
This deviation is largely caused by spectral hardening, which is
parameterized by the hardening factor $f = T_{\rm in}/T_{\rm eff}$,
where $T_{\rm in}$ and $T_{\rm eff}$ are respectively the color
temperature and the effective temperature (Shimura \& Takahara 1995;
Davis et al.\ 2005).  The hardening factor $f$ increases with luminosity
so that when luminosity is plotted versus $T_{\rm eff}$, one obtains a
relation that is much closer to $L \propto T^4$ (Davis et al.\ 2006;
McClintock et al.\ 2007).

The significance of the constancy of the inner disk radius was noted
early on; for example, Tanaka \& Lewin (1995) commented that the
constancy of this inner radius suggests that it is related to the
innermost stable circular orbit (ISCO) predicted by general relativity.
This identification of the inner disk radius with the ISCO is the
foundation for the measurements of black hole spin that have been
obtained recently by two independent methods: fitting the spectrum of
the X-ray continuum (Shafee et al.\ 2006; McClintock et al.\ 2006; Liu
et al.\ 2008; Gou et al.\ 2009) and modeling the profile of the Fe K
line (Brenneman \& Reynolds 2006; Miller et al.\
2008; Reis et al.\ 2008).

\subsection{Impulsive Power Law Flares}

The response of the disk parameters to intense flares has been observed
for several black hole binaries (e.g., Sobczak et al.\ 1999, 2000b) and
has been attributed to increased spectral hardening or Compton
upscattering of soft disk photons; the actual physical radius may remain
fairly constant.  For a thorough discussion see Sobczak et al.\ (2000b).

The most remarkable power-law flare observed for a black hole binary
occurred early in the outburst cycle of XTE J1550--564.  The flare
reached a peak intensity of 6.8 Crab (2--10 keV) on 1998 September
19--20 (Sobczak et al.\ 2000b).  Soon after this strong flare,
superluminal radio jets ($>2c$) were reported (Hannikainen et al.\
2001).  A few years later, large-scale ($\sim 20$ arcsec) radio and
X-ray jets were discovered, which were most likely ejected during the
September 1998 event (Corbel et al.\ 2002; Tomsick et al.\ 2003; Kaaret
et al.\ 2003).  More recently, a similar pair of large-scale, X-ray jets
have been observed for H1743 (Corbel et al.\ 2005).  By analogy with the
behavior of XTE J1550--564, it is reasonable to suppose that these jets
are connected with the impulsive flare events described in \S3.4.2.

\subsection{Radio/X-ray Connections}

It is unreasonable to expect a point-by-point correlation between the
X-ray and radio data shown in Figure 13 because the higher time
resolution ASM light curve (Fig.\ 3b) shows strong X-ray variability on
a time scale as short as minutes, and the X-ray and radio observations
were typically separated by $\gtrsim$ hours (compare Tables A2 \& A3).
For example, for the tripling of the radio intensity that occurred on
Day 18.5 (Fig.\ 15; Table A3), the proximate X-ray observation did not
commence for over 2 hours (Table A2).  Similarly, the radio observation
nearest the intense power-law X-ray flare that occurred on Day 46.9
(Table A2) occurred 10.7 hours earlier.  Similarly, given the sparseness
and asynchronous quality of the X-ray/radio data, it is not possible to
make useful comments on the relationship between radio flaring and state
transitions (e.g., see Fender et al.\ 2004).

Despite the limitations of the data, all three of the X-ray quantities
shown in Figure 14$b$--$d$ do appear to be generally correlated with the
radio data.  Arguably it is the break in the power-law spectrum at $\sim
15$ keV ($\Gamma_{\rm HI} - \Gamma$) that best correlates with the radio
data.


\section{Conclusions}

We have performed a spectral and timing analysis of 170 {\it RXTE}
observations that cover the complete 2003 outburst cycle of H 1743-322.
The PCA and HEXTE spectra, which were fitted simultaneously, were
decomposed into disk-blackbody and power-law components.  For about a
quarter of the spectra, a much better fit was achieved when a break in
the spectral index was included in the power-law component.  Likewise, a
Gaussian Fe-K emission line component was included whenever it improved
the fit significantly.  No thermal emission was detected in 16 spectra,
and the disk black-body component was not used in fitting these spectra.
The model gave good fits to all but three of the 170 spectra.  Power
density spectra were computed and the level of the power continuum was
determined for all 170 observations.  QPOs in the range 1--15 Hz were
detected for 87 of the data sets, and the frequency, intensity and
quality factor Q of the dominant QPO were obtained.  All of the products
of these spectral and timing analyses are contained in Tables A1 \& A2,
and most of the results are summarized in Figures 4 \& 5, which present
a detailed look at the spectral and temporal evolution of the 2003
outburst.

Using the results contained in Tables A1 \& A2, we classified the state
of the source according to the criteria summarized in Table~1
(McClintock \& Remillard 2006; RM06), which are based on the power-law
spectral index, the relative strengths of the power-law and disk
components, the presence or absence of QPOs, and the strength of the
power density continuum.  These state classifications, which are a key
ingredient of this work, are displayed in Figures 3--5, 8--11, \&
13--14, listed in Table A2, and referred to throughout the text.

The X-ray properties of H1743 and XTE J1550--564 are very similar,
including the general character of their X-ray light curves, the
occasional appearance of impulsive power-law flares (Fig.\ 12), and
their long and slow decline in the thermal state (Figs.\ 3 \& 4), during
which their disk luminosities are approximately $\propto T_{\rm
in}^4$~(Fig.\ 7).  Additional notable similarities are the nearly
identical and tight correlations that exist between the low-luminosity
disk flux, the QPO frequency, and the 0.1--1 Hz continuum power.  These
results, which are displayed in Figures 8--11, illustrate that the
transition between the hard and SPL states can be remarkably orderly and
stereotypical.  As shown in Figure~5, the intermediate H:SPL state is
distinguished by its strong QPOs and by its high efficiency for
producing Fe-line flux.

We present VLA data at six frequencies that extend throughout the entire
outburst cycle at 4.860 GHz and 8.460 GHz.  The radio source was
observed to be active only during the first $\approx 100$ days (Fig.\
13).  Finally, we present optical/NIR data and finding charts (Figs.\ 15
\& 16) for the optical counterpart in both outburst and quiescence.

\acknowledgments

This work was supported in part by NASA grants NNG05GB31G and
NNX08AJ55G.  MAPT was supported by NASA LTSA grant NAG-5-10889 and DS
acknowledges the support of a Smithsonian Astrophysical Observatory Clay
fellowship.  We thank Jean Swank and her staff for support and for the
execution of the {\it RXTE} observations, David Kaplan for obtaining the
2003 Magellan/MagIC images, Jack Steiner for advice on statistics, and
an anonymous referee whose criticisms were quite helpful.

\appendix

\section{Results of X-ray and Radio Analyses in Tabular Form}

Tables A1 \& A2 contain all of the X-ray spectral parameters and fluxes
that are displayed in Figures 4 \& 5 for the 170 observations.  The
tables jointly also relate the Day Number (used in Figs.\ 1--5 and
Figs.\ 13 \& 14) to the UT date and MJD, and they assign an index number
to each observation.  For each observation, Table A1 also provides the
following additional information that is not presented in the figures:
the break energy $E_{\rm break}$, the central energy and FWHM of the Fe
line, and the reduced chi-square and number of degrees of freedom.
Likewise, Table A2 contains two quantities that are not displayed in the
figures: the exposure time for each observation and the ratio of the
disk flux to the total flux (2--20 keV).

All of the radio flux density data plotted in Figures~13~\&~14 versus
Day Number are contained in Tables A3 \& A4, respectively.  These tables
also give the correspondence between Day Number and the UT date and MJD.
Table A5 gives 25 values of the radio spectral index that were
determined during the first 115 days of the outburst cycle (see Fig.\
13$d$).

\begin{deluxetable}{ccccccccccccc}
\tabletypesize{\scriptsize}
\rotate
\tablewidth{0pt}
\tablenum{A1}
\tablecaption{Spectral Parameters for H1743-322}
\tablehead{
 \colhead{Obs.} & 
 \colhead{ } & 
 \colhead{ } &
 \colhead{$T_{\rm in}$} & 
 \colhead{$R^*_{\rm in}$\tablenotemark{c}} & 
 \colhead{ } & 
 \colhead{ } & 
 \colhead{$E_{\rm break}$} &
 \colhead{Power-Law} & 
 \colhead{$E_{\rm line}$\tablenotemark{g}} & 
 \colhead{FWHM\tablenotemark{g}} &
 \colhead{Line} & 
 \colhead{ }
\cr
 \colhead{No.} & 
 \colhead{Day\tablenotemark{a}} & 
 \colhead{MJD\tablenotemark{b}} & 
 \colhead{(keV)} & 
 \colhead{(km)} & 
 \colhead{$\Gamma$\tablenotemark{d}} & 
 \colhead{$\Gamma_{\rm HI}$\tablenotemark{e}} &
 \colhead{(keV)} &
 \colhead{Norm.\tablenotemark{f}} & 
 \colhead{(keV)} & 
 \colhead{(keV)} & 
 \colhead{Flux\tablenotemark{g}} & 
 \colhead{$\chi_{\nu}^2$/dof\tablenotemark{h}}
\cr
 \colhead{(1)} &
 \colhead{(2)} &
 \colhead{(3)} &
 \colhead{(4)} &
 \colhead{(5)} &
 \colhead{(6)} &
 \colhead{(7)} &
 \colhead{(8)} &
 \colhead{(9)} &
 \colhead{(10)} &
 \colhead{(11)} &
 \colhead{(12)} &
 \colhead{(13)}
}
\startdata
 
1  &   7.8&  52726.850&   $1.776\pm0.073$&  $ 1.21\pm0.07$&   $1.24\pm0.03$&  $1.92\pm0.03$&  $22.7\pm0.6$&  $ 0.09\pm0.01$&   \nodata&      \nodata&      \nodata&        1.65/87\\

2  &  10.8&  52729.809&   \nodata&          \nodata&          $1.56\pm0.01$&  $2.19\pm0.02$&  $16.6\pm0.4$&  $ 0.46\pm0.01$&   $6.4\pm0.2$&  $0.5\pm0.2$&  $0.30\pm0.08$&  1.70/86\tablenotemark{i}\\

3  &  14.8&  52733.762&   \nodata&          \nodata&          $2.27\pm0.01$&  $2.89\pm0.02$&  $17.1\pm0.4$&  $ 3.72\pm0.04$&   $6.5\pm0.2$&  $0.5\pm0.2$&  $0.52\pm0.14$&  1.08/86\\

4  &  16.7&  52735.734&   $1.265\pm0.021$&  $ 8.30\pm0.44$&   $2.23\pm0.04$&  $3.00\pm0.02$&  $15.1\pm0.3$&  $ 3.52\pm0.32$&   \nodata&      \nodata&      \nodata&        0.93/87\\
 
5  &  18.6&  52737.580&   \nodata&          \nodata&          $2.28\pm0.01$&  $3.01\pm0.02$&  $15.7\pm0.2$&  $ 6.82\pm0.07$&   $6.5\pm0.1$&  $0.5\pm0.1$&  $1.28\pm0.19$&  0.91/86\\
 
6  &  20.7&  52739.678&   \nodata&          \nodata&          $2.52\pm0.02$&  $3.18\pm0.02$&  $16.3\pm0.3$&  $11.77\pm0.12$&   $6.4\pm0.2$&  $0.3\pm0.2$&  $0.83\pm0.19$&  0.82/86\\
\enddata
\scriptsize
\tablenotetext{a}{Number of days since discovery on 2003 March 21 = MJD 52719 (Revnivtsev et al. 2003); for calendar dates, see Table A3.}
\tablenotetext{b}{Start time of observation. MJD = JD - 2,400,000.5; for calendar dates, see Table A3.}
\tablenotetext{c}{$R^*_{\rm in}=R_{\rm in} (\cos \theta)^{1/2} / (D/10~kpc)$,
where $\theta$ is the inclination angle and $D$ is the distance to the source in kpc.}
\tablenotetext{d}{Photon index for energy band 3 keV to $E_{\rm break}$ for broken power-law fits and $E_{\rm break}$ to 200 keV index for simple power-law fits.}
\tablenotetext{e}{Photon index for energy band $E_{\rm break}$ to 200 keV for broken power-law fits.}
\tablenotetext{f}{Photons s$^{-1}$ cm$^{-2}$ keV$^{-1}$ at 1~keV.}
\tablenotetext{g}{Three Gaussian line parameters: central line energy, full width at half maximum intensity, and Fe K flux in photons~s$^{-1}$~cm$^{-2}$.}
\tablenotetext{h}{Reduced $\chi^2$; number of degrees of freedom.}
\tablenotetext{i}{Fitted using data from HEXTE Cluster B because data from Cluster A saturated the telemetry.}
\tablenotetext{j}{HEXTE fitting range restricted to $E = 32-200$ keV (includes HEXTE-A channels 17--56).}
\tablenotetext{k}{PCA background appears to be overestimated in standard processing, which probably contributes to poor fit.}

\end{deluxetable}

\clearpage

\begin{deluxetable}{cccccccccccccc}
\tabletypesize{\scriptsize} 
\rotate
\tablewidth{0pt}
\tablenum{A2}
\tablecaption{Unabsorbed Flux Data\tablenotemark{a}, QPO Data, and State Classifications for H1743-322}
\tablehead{
 \colhead{} & 
 \colhead{} & 
 \colhead{} & 
 \colhead{Obs.} & 
 \colhead{Total} & 
 \colhead{Disk} &
 \colhead{Power Law} & 
 \colhead{} & 
 \colhead{Power Law} & 
 \colhead{QPO} & 
 \colhead{QPO} & 
 \colhead{QPO} &
 \colhead{Cont.}&
 \colhead{}
\cr
 \colhead{Obs.} & 
 \colhead{} & 
 \colhead{UT 2003} & 
 \colhead{Time} & 
 \colhead{2--20 keV} & 
 \colhead{2--20 keV} &
 \colhead{2--20 keV} & 
 \colhead{2--20 keV} & 
 \colhead{20--100 keV} & 
 \colhead{$\nu$\tablenotemark{c}} & 
 \colhead{Amplitude\tablenotemark{d}} & 
 \colhead{$Q$\tablenotemark{e}} & 
 \colhead{$r$\tablenotemark{f}}&
 \colhead{} 
\cr
 \colhead{No.} & 
 \colhead{Day\tablenotemark{b}} & 
 \colhead{(mmdd)} & 
 \colhead{(s)} & 
 \colhead{($10^{-8}$ cgs)} & 
 \colhead{($10^{-8}$ cgs)} & 
 \colhead{($10^{-8}$ cgs)} & 
 \colhead{Disk/Total} & 
 \colhead{($10^{-9}$ cgs)} & 
 \colhead{(Hz)} & 
 \colhead{(\% rms)} & 
 \colhead{}& 
 \colhead{(rms)}&
 \colhead{State\tablenotemark{g}}
 }
\startdata 
1 &   7.8 &  0328 &   3472 & $0.175_{-0.001}^{+0.001}$ &  $ 0.022_{-0.001}^{+0.001}$ &  $ 0.153_{-0.001}^{+0.001}$ &  $0.126_{-0.006}^{+0.006}$ & $2.623_{-0.029}^{+0.022}$ &     0.063 & $3.63\pm0.45$ &   $31.50\pm11.07$ &  0.242 &? \\
   
2 &  10.8 &  0331 &   3488 & $0.396_{-0.002}^{+0.001}$ &          \nodata            &  $ 0.396_{-0.001}^{+0.001}$ &        \nodata             & $3.396_{-0.023}^{+0.030}$ &     0.159\tablenotemark{h} & $13.72\pm0.58$ &  $3.79\pm0.27$ &  0.262 &H\tablenotemark{i} \\
   
3 &  14.8 &  0404 &   3472 & $0.846_{-0.005}^{+0.003}$ &          \nodata            &  $ 0.846_{-0.001}^{+0.001}$ &        \nodata             & $2.060_{-0.034}^{+0.026}$ &     2.23\tablenotemark{j} &  $12.53\pm0.23$ &  $11.64\pm0.62$ & 0.206 &H:SPL \\
   
4 &  16.7 &  0406 &   3488 & $1.067_{-0.004}^{+0.003}$ &  $ 0.216_{-0.003}^{+0.001}$ &  $ 0.851_{-0.002}^{+0.002}$ &  $0.202_{-0.003}^{+0.002}$ & $1.824_{-0.034}^{+0.017}$ &     3.75\tablenotemark{j} &  $12.32\pm0.20$ &  $11.36\pm0.35$ & 0.173 &H:SPL \\
   
5 &  18.6 &  0408 &   3568 & $1.518_{-0.006}^{+0.003}$ &           \nodata           &  $ 1.518_{-0.002}^{+0.002}$ &        \nodata             & $3.148_{-0.026}^{+0.022}$ &     2.08\tablenotemark{j} &  $12.23\pm0.27$ &  $11.56\pm0.64$ &  0.197 &H:SPL \\
\enddata
\end{deluxetable}

\clearpage

\begin{deluxetable}{ccccccccccc}
\tabletypesize{\scriptsize} 
\tablewidth{0pt}
\tablenum{A3}
\tablecaption{VLA Radio Flux Densities for H1743--322: $\nu \leq 8.640$ GHz}
\tablehead{  
 \colhead{UT 2003} & \colhead{} & \colhead{} & \colhead{} & \colhead{1.425 GHz} &
 \colhead{} & \colhead{} & \colhead{4.860 GHz} & \colhead{} & \colhead{} & \colhead{8.460 GHz} \cr
 \colhead{(mmdd)} & \colhead{MJD\tablenotemark{a}} & \colhead{Day\tablenotemark{b}} & 
 \colhead{VLA\tablenotemark{c}} & \colhead{$S_{\nu}$(mJy)} & \colhead{MJD} & \colhead{Day} &
 \colhead{$S_{\nu}$(mJy)} & \colhead{MJD} & \colhead{Day} & \colhead{$S_{\nu}$(mJy)}
}
\startdata
0330&  \nodata&     \nodata&  D&    \nodata&         52728.6016&   9.6&     $ 3.39\pm0.15$&  52728.5938&  9.6&     $ 4.57\pm0.12$ \\  
0401&  52730.5586&  11.6&     D&    $8.19\pm1.53$&   52730.5547&  11.6&     $ 5.78\pm0.18$&  52730.5469&  11.6&    $ 6.45\pm0.12$ \\  
0403&  \nodata&     \nodata&  D&    \nodata&         52732.5195&  13.5&     $15.31\pm0.20$&  52732.5469&  13.6&    $16.99\pm0.12$ \\  
0404&  \nodata&     \nodata&  D&    \nodata&         52733.5195&  14.5&     $20.08\pm0.24$&  52733.5156&  14.5&    $18.68\pm0.31$ \\   
0404&  \nodata&     \nodata&  D&    \nodata&         \nodata&     \nodata&  \nodata&         52733.5469&  14.6&    $21.81\pm0.13$ \\  
0406&  \nodata&     \nodata&  D&    \nodata&         52735.4062&  16.4&     $20.27\pm0.26$&  52735.4219&  16.4&    $19.62\pm0.16$ \\
0406&  \nodata&     \nodata&  D&    \nodata&         52735.4375&  16.4&     $20.47\pm0.18$&  52735.4414&  16.4&    $19.58\pm0.13$ \\
0406&  \nodata&     \nodata&  D&    \nodata&         52735.4688&  16.5&     $20.23\pm0.22$&  52735.4648&  16.5&    $19.43\pm0.16$ \\
\enddata 
\scriptsize   
\tablenotetext{a} {Modified Julian Day = JD -- 2,400,000.5.}
\tablenotetext{b} {Number of days since discovery on 2003 March 21 = MJD 52719 (Revnivtsev et al. 2003).}
\tablenotetext{c} {Configuration of the VLA antennas.}
\tablenotetext{d} {Upper limits correspond to the flux density measured at the nominal position of the center of the beam plus three times the rms noise.}          

\end{deluxetable}

\clearpage

\begin{deluxetable}{cccccccccc}
\tabletypesize{\scriptsize} 
\tablewidth{0pt}
\tablenum{A4}
\tablecaption{VLA Radio Flux Densities for H1743--322: $\nu \geq 14.940$ GHz}
\tablehead{  
 \colhead{UT 2003} & \colhead{} & \colhead{} & \colhead{14.940 GHz} & 
 \colhead{} & \colhead{} & \colhead{22.460 GHz} & \colhead{} & \colhead{} & \colhead{43.340 GHz} \cr
 \colhead{(mmdd)} & \colhead{MJD\tablenotemark{a}} & \colhead{Day\tablenotemark{b}} & 
 \colhead{$S_{\nu}$(mJy)} & \colhead{MJD} & \colhead{Day} &
 \colhead{$S_{\nu}$(mJy)} & \colhead{MJD} & \colhead{Day} & \colhead{$S_{\nu}$(mJy)}
}
\startdata
0401&   52730.5430&  11.5&     $6.81\pm0.51$&    \nodata&     \nodata&  \nodata&          \nodata&     \nodata&  \nodata       \\
0403&   52732.5391&  13.5&     $19.20\pm0.33$&   52732.5352&  13.5&     $21.32\pm0.23$&   52732.5234&  13.5&     $21.54\pm0.87$ \\
0404&   52733.5430&  14.5&     $26.57\pm0.40$&   52733.5391&  14.5&     $26.36\pm0.26$&   52733.5273&  14.5&     $27.55\pm1.01$ \\
0406&   52735.4570&  16.5&     $20.15\pm0.45$&   52735.4531&  16.5&     $22.69\pm0.46$&   52735.4453&  16.4&     $16.46\pm1.26$ \\
0408&   52737.5000&  18.5&     $42.62\pm0.44$&   52737.5078&  18.5&     $51.01\pm0.25$&   52737.4961&  18.5&     $41.11\pm0.73$ \\
0408&   52737.5156&  18.5&     $55.21\pm0.32$&   \nodata&     \nodata&  \nodata&          \nodata&     \nodata&  \nodata       \\
0409&   52738.4805&  19.5&     $19.58\pm0.31$&   52738.4766&  19.5&     $21.13\pm0.17$&   52738.4766&  19.5&     $19.29\pm0.46$ \\
0404&   \nodata&     \nodata&  \nodata&          \nodata&     \nodata&  \nodata&          52738.4727&  19.5&     $19.37\pm0.62$ \\
\enddata 
\scriptsize   
\tablenotetext{a} {Truncated Modified Julian Day = MJD - 52000.} 
\tablenotetext{b} {Number of days since discovery on 2003 March 21 = MJD 52719 (Revnivtsev et al. 2003).}  
\tablenotetext{c} {Upper limits correspond to the flux density measured at the nominal position of the center of the beam plus three times the rms noise.}
\end{deluxetable}

\clearpage

\begin{deluxetable}{cccc}
\small
\tablewidth{0pt}  
\tablenum{A5}
\tablecaption{Radio Spectral Index}

\tablehead{  
\colhead{UT 2003} & 
\colhead{} & \colhead{} & 
\colhead{Spectral Index} \cr
\colhead{(mmdd)} & 
\colhead{MJD\tablenotemark{a}} & 
\colhead{Day\tablenotemark{b}} & 
\colhead{$\alpha$\tablenotemark{c}}
}
\startdata

0330&  52728.598&    9.6&   $ 0.54\pm0.09$ \\
0401&  52730.551&   11.6&   $ 0.20\pm0.07$ \\  
0404&  52733.518&   14.5&   $-0.13\pm0.04$ \\
0406&  52735.414&   16.4&   $-0.06\pm0.03$ \\
0406&  52735.439&   16.4&   $-0.08\pm0.02$ \\
0406&  52735.467&   16.5&   $-0.07\pm0.02$ \\
0408&  52737.490&   18.5&   $-0.62\pm0.01$ \\
0409&  52738.459&   19.5&   $-0.23\pm0.02$ \\
0411&  52740.551&   21.6&   $-0.79\pm0.02$ \\
0412&  52741.563&   22.6&   $-0.91\pm0.04$ \\
0413&  52742.520&   23.5&   $-0.23\pm0.11$ \\
0416&  52745.436&   26.4&   $-0.20\pm0.01$ \\
0420&  52749.602&   30.6&   $-0.50\pm0.05$ \\
0421&  52750.598&   31.6&   $-0.18\pm0.05$ \\
0423&  52752.535&   33.5&   $-0.11\pm0.05$ \\
0429&  52758.418&   39.4&   $-0.43\pm0.04$ \\
0506&  52765.430&   46.4&   $-0.39\pm0.03$ \\
0508&  52767.520&   48.5&   $-0.19\pm0.02$ \\
0509&  52768.498&   49.5&   $-0.25\pm0.02$ \\
0510&  52769.512&   50.5&   $-0.30\pm0.02$ \\
0527&  52786.352&   67.3&   $-0.64\pm0.05$ \\
0610&  52800.246&   81.2&   $-1.45\pm0.24$ \\
0619&  52809.219&   90.2&   $-0.35\pm0.12$ \\
0709&  52829.344&  110.3&   $-1.89\pm0.26$ \\
0713&  52833.217&  114.2&   $-0.90\pm0.55$ \\

\enddata
\footnotesize   
\tablenotetext{a} {Modified Julian Day = JD -- 2,400,000.5.}
\tablenotetext{b} {Number of days since discovery on 2003 March 21 = 
MJD~52719 (Revnivtsev et al. 2003).}
\tablenotetext{c} {Radio spectral index $\alpha$ ($S_{\rm \nu} \propto
  \nu^{\alpha}$), which is based on the 4.860 GHz and 8.460 GHz data.}

\end{deluxetable}

\newpage

\figcaption[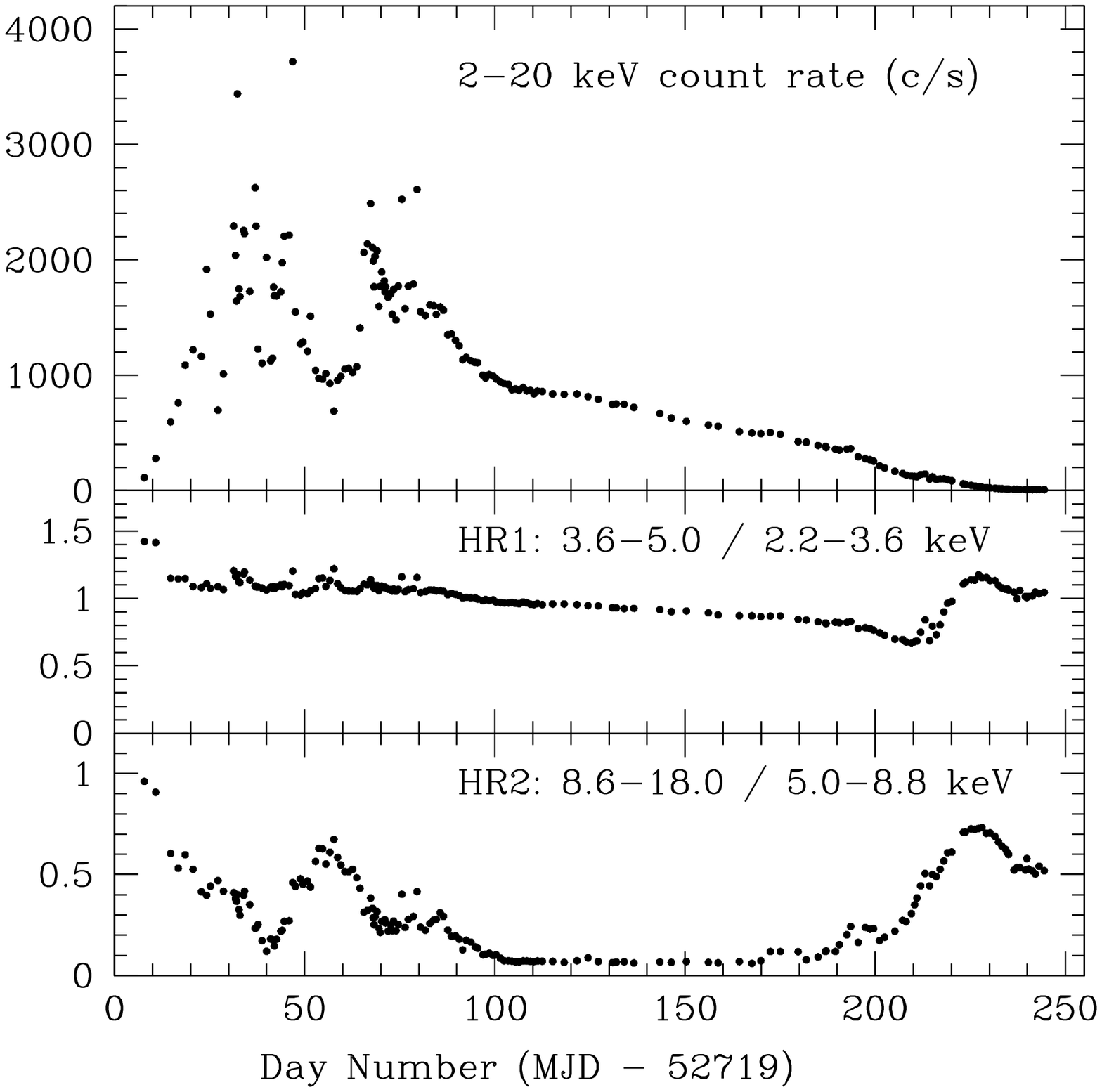] {{\it RXTE} PCA count rates and hardness ratios for
  the energy intervals indicated.  These data, from which the background
  rates have been subtracted (\S2.2), were obtained using Proportional
  Counter Unit 2.  The Day Number is the time in days since the
  discovery of the outburst on 2003 March 21 (= MJD 52719).}

\figcaption[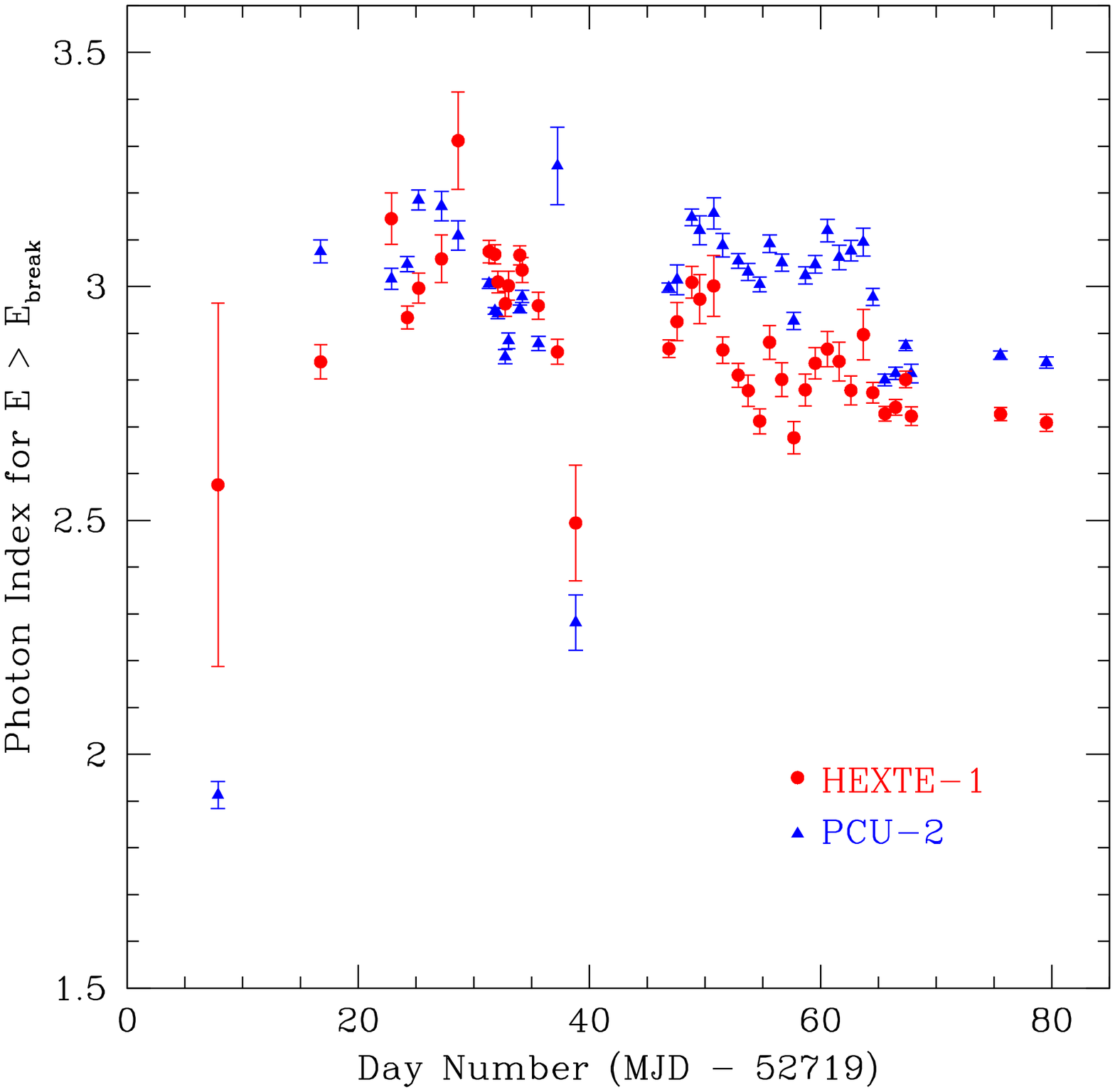] {The high-energy photon index ($E > E_{\rm break}$)
  was fitted independently for the PCU-2 and the HEXTE-A detectors, and
  the results are compared here.  For each observation, an
  inverse-variance-weighted average of the two indices was computed to
  give the single photon index, \GHI~(see Fig.\ 4$d$ and Table A1).}

\figcaption[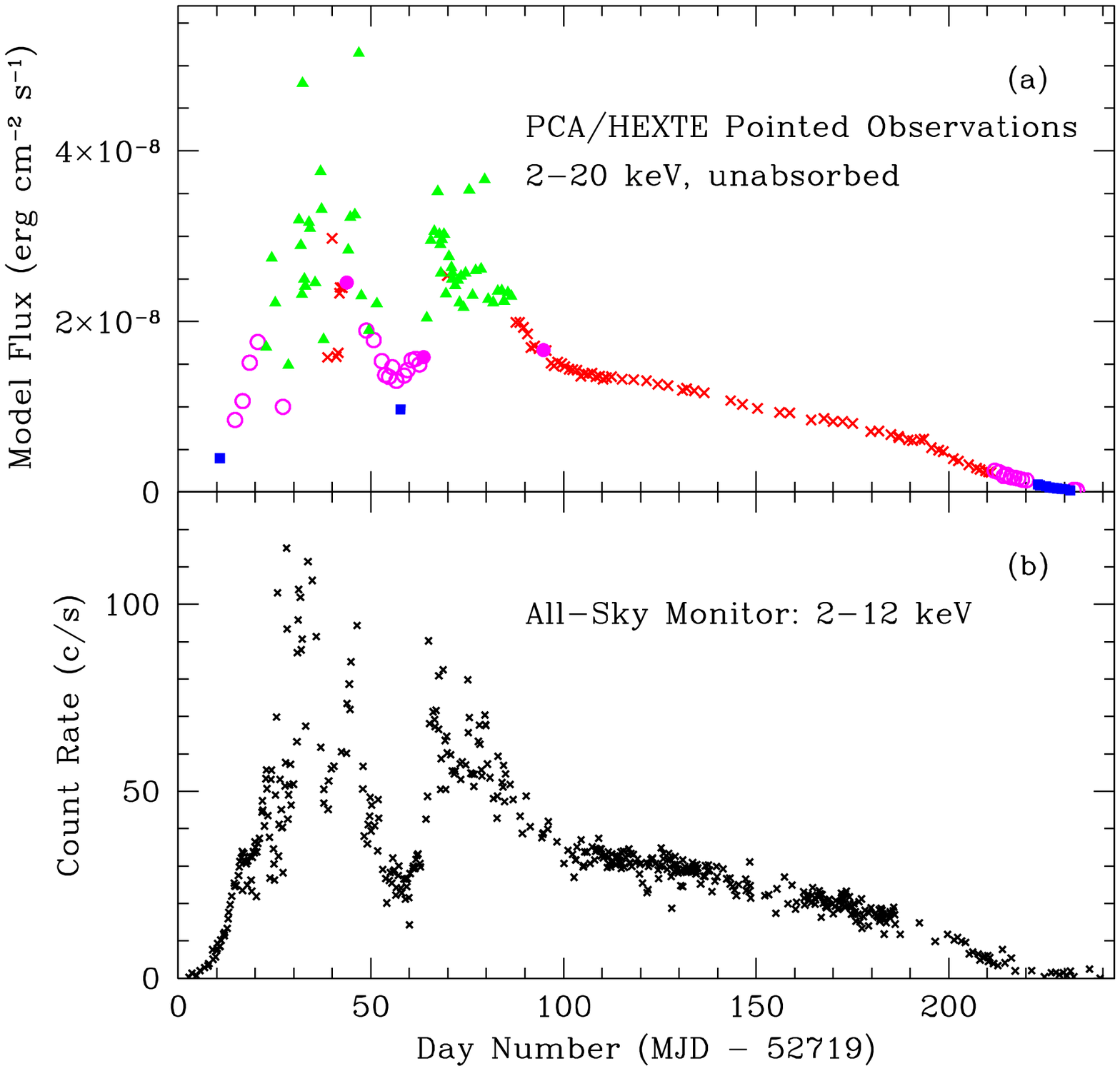] {The evolution of H1743 during its 2003 outburst.
  ($a$) The total 2--20 keV unabsorbed model flux given in the fifth
  column of Table~A2. The symbol type denotes the X-ray state (see
  text): thermal (red cross); hard (blue square); steep power-law (green
  triangle); H:SPL intermediate state (magenta open circle); and TD:SPL
  intermediate state (magenta filled circle). ($a$) The ASM light curve;
  the data for individual dwells, up to several per day, are plotted.}

\figcaption[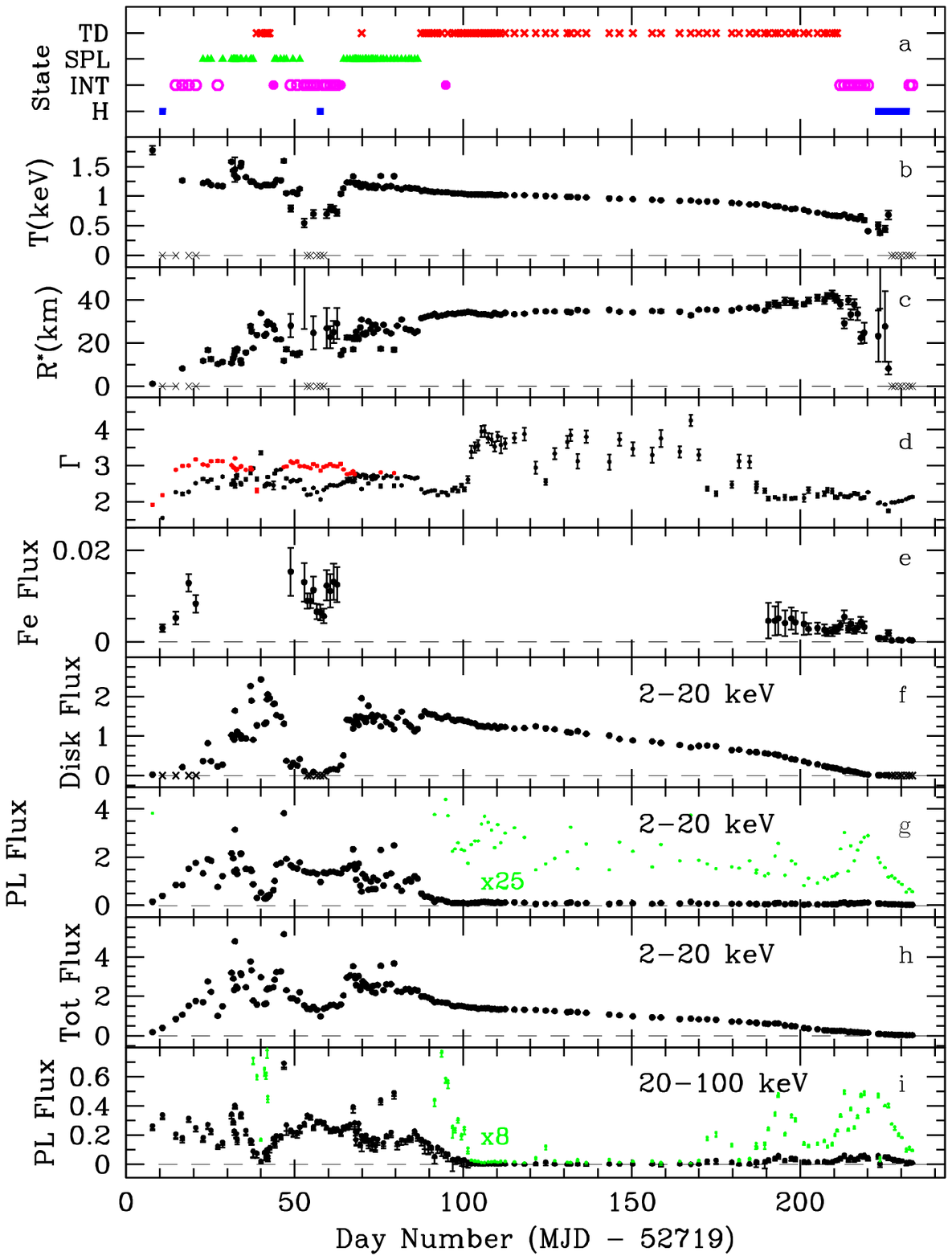] {Spectral parameters and fluxes for 170 observations
  versus Day Number.  All the data presented here are contained in
  Tables A1 \& A2.  The state of the source is indicated in ($a$).  The
  spectral parameters are ($b$) the color temperature of the accretion
  disk \Tin~in keV; ($c$) the inner disk radius $R^*$(km)$ \equiv R_{\rm
    in}($cos$~i)^{1/2}/(D/10$ kpc), where $i$ is the inclination angle
  and $D$ is the distance to the source in kpc; and ($d$) the power-law
  photon indices, where the black data points are for $\Gamma$~($E <
  E_{\rm break}$) and the red points are for \GHI~($E > E_{\rm break}$);
  see columns~6 \& 7 in Table~A1.  Note the exceptional observation on
  Day 38.8 for which the power-law component hardens at high energies,
  i.e., $\Gamma > \Gamma_{\rm HI}$.  The unabsorbed fluxes shown in
  units of $10^{-8}$~\flux~are for ($e$) the Fe-K line; ($f$) the disk
  (2--20 keV); ($g$) the power law (2--20 keV); ($h$) the total (2--20
  keV); and ($i$) the high-energy power law (20--100 keV).  All the data
  are plotted with error bars, although they are usually too small to be
  apparent.  In order to better display the power-law fluxes in panels
  $g$ \& $i$ when they are faint, these quantities have been multiplied
  by the factor indicated in their respective panels and replotted as
  small, green symbols.}

\figcaption[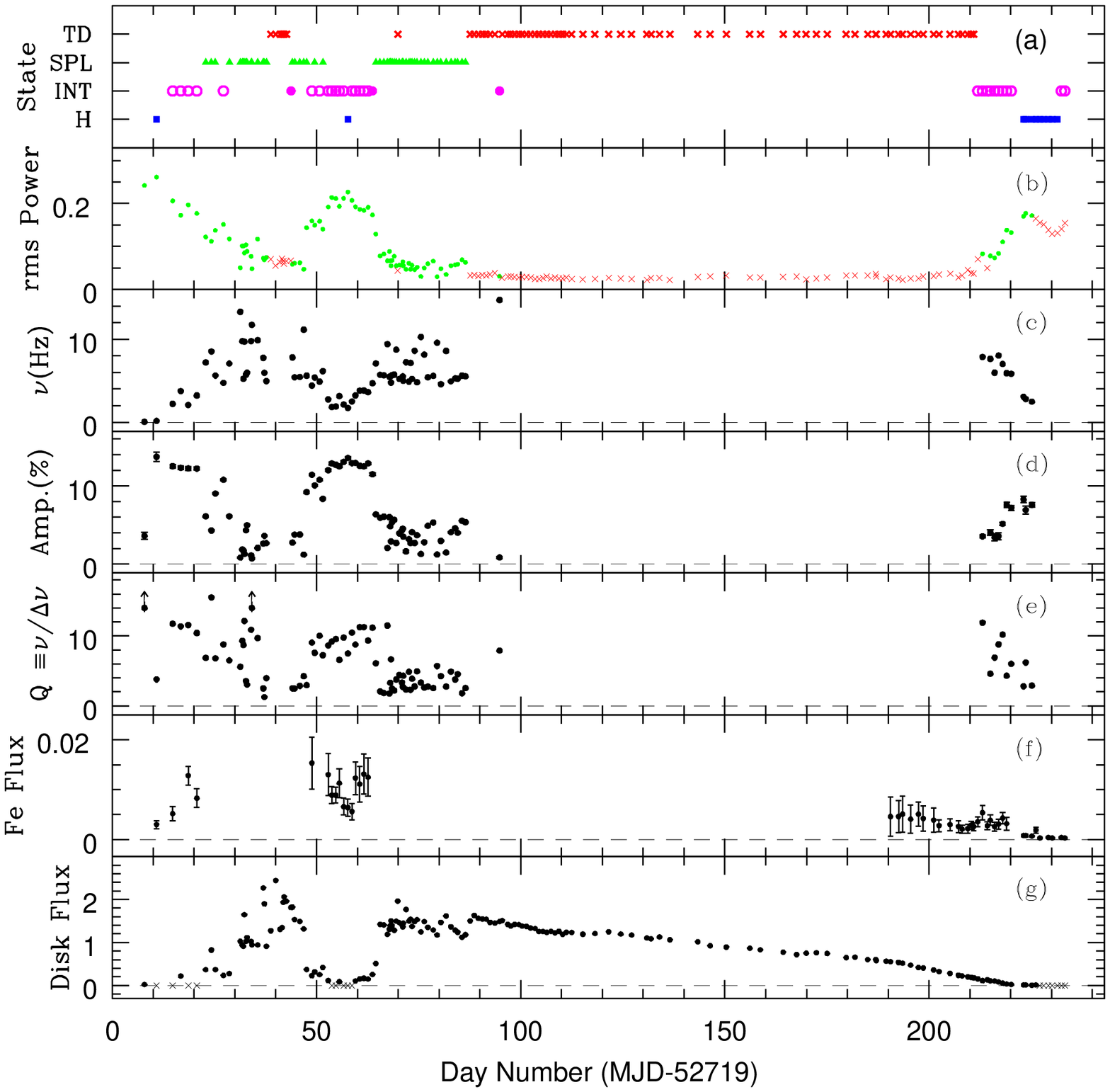] {Timing data based on a power density spectrum
  analysis plotted versus Day Number.  The state classification is shown
  in ($a$), and the 0.1--10 Hz rms continuum power in the PDS is shown
  in ($b$).  For the dominant low frequency QPO, the following three
  quantities are plotted: ($c$) central QPO frequency; ($d$) amplitude
  (\% rms); and ($e$) quality factor $Q$.  The gaps in these plots
  correspond to the non-detection of QPOs.  For reference, the Fe K-line
  and accretion-disk fluxes ($f$ \& $g$, respectively) are shown
  replotted from Figure 4.}

\figcaption[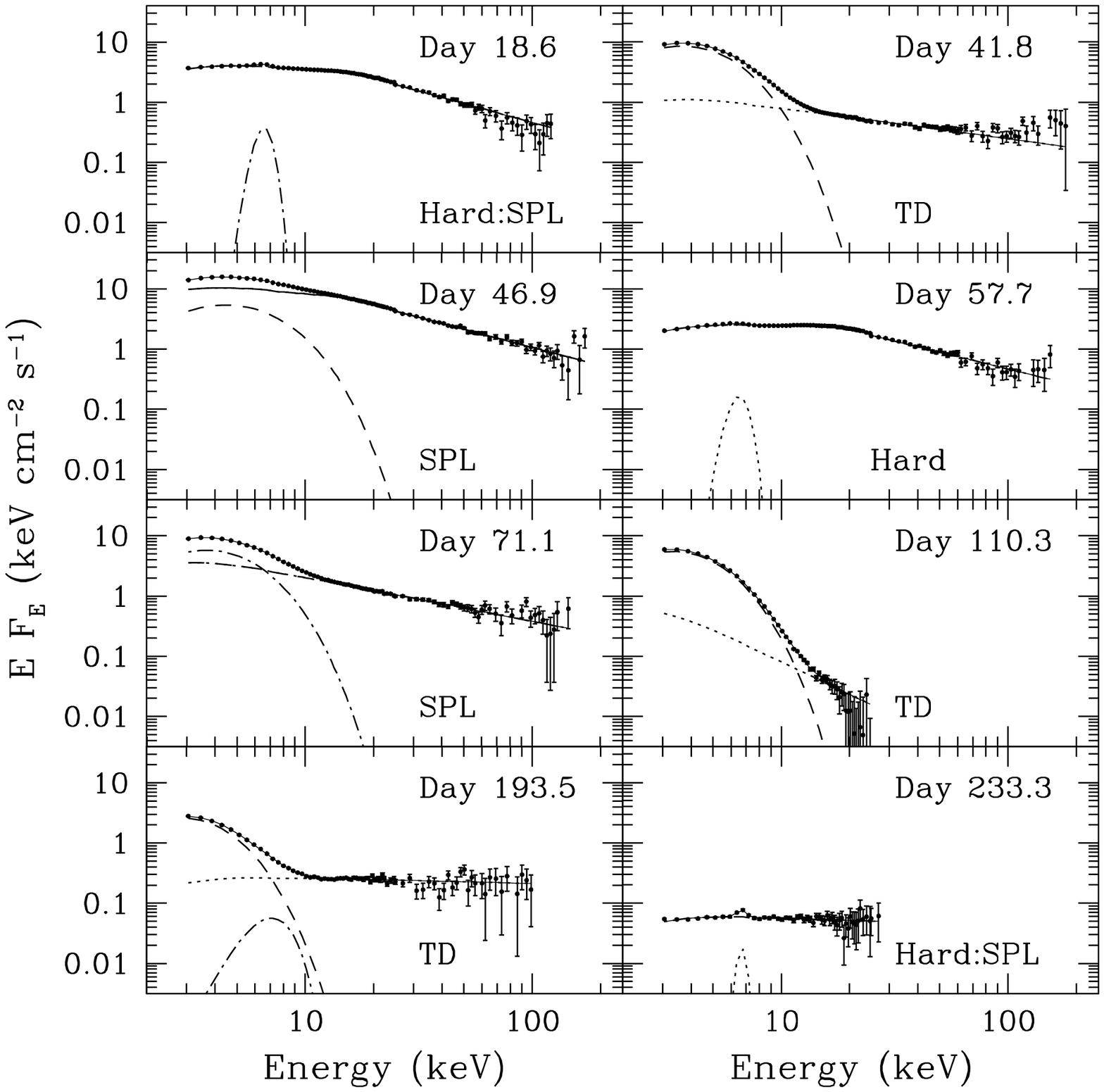] {Sample unabsorbed and unfolded energy spectra
  identified by Day Number and typed by X-ray state.  For each panel,
  the solid line, which is largely obscured by the data, is the best-fit
  model spectrum.  The power-law, disk and Fe-line components are shown
  as dotted, dashed and dot-dashed lines, respectively.  Day 18.6:
  spectrum from the rising phase dominated by the power-law; note break
  near 15 keV.  Day 41.8: an interlude of thermal-state behavior during
  the flaring state.  Day 46.9: spectrum of the strongest power-law
  flare; the power-law is dominant, but note the hot ($kT = 1.60$ keV)
  and luminous disk component.  Day 57.7: this brief excursion to the
  hard state during the course of an outburst cycle is unusual behavior
  for an X-ray transient.  Day 110.3: representative spectrum obtained
  well into the thermal-state decay of the disk; note the faint, steep
  power-law component.  Day 193.5: representative spectrum during the
  later stage of the thermal-state decay of the disk; note the hard,
  prominent power-law component during this period.  Day 233.3: the
  final spectrum with $\Gamma = 2.14 \pm 0.02$ (Table A1) is essentially
  a hard-state spectrum (see footnote $g$ in Table~A3.}

\figcaption[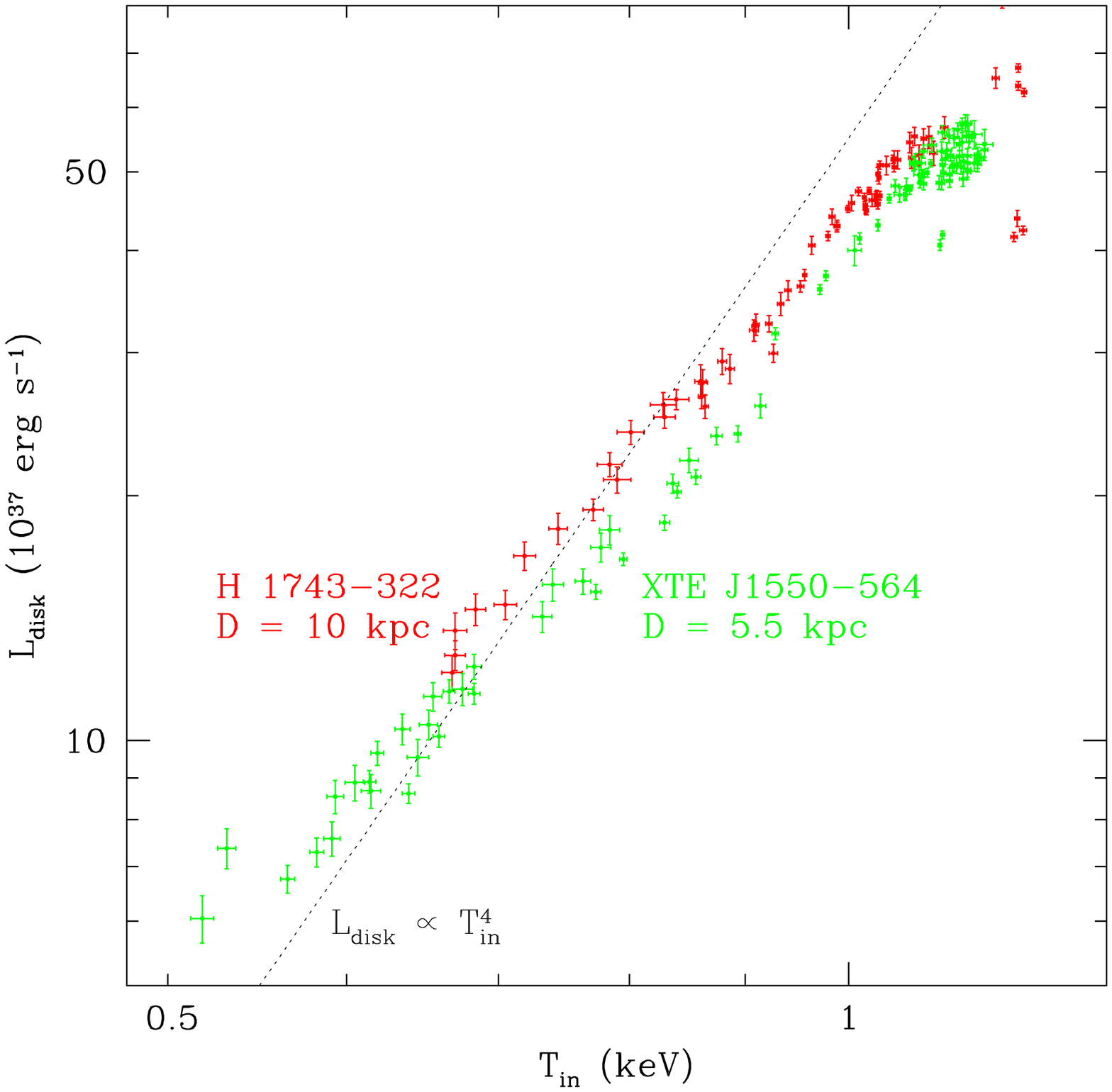] {Bolometric luminosities versus disk temperature for
  observations made in the thermal state.  For H1743 the eight data
  points at the highest temperatures ($T_{\rm in} \approx 1.2$ keV) were
  made early in the outburst cycle, near Day 42, when the flaring
  temporarily subsided; note that some of these thermal-state data fall
  well below an extrapolation of the lower-temperature data.  For
  reference, the dotted line indicates the slope for $L_{\rm disk}
  \propto T_{\rm in}^4$.}

\figcaption[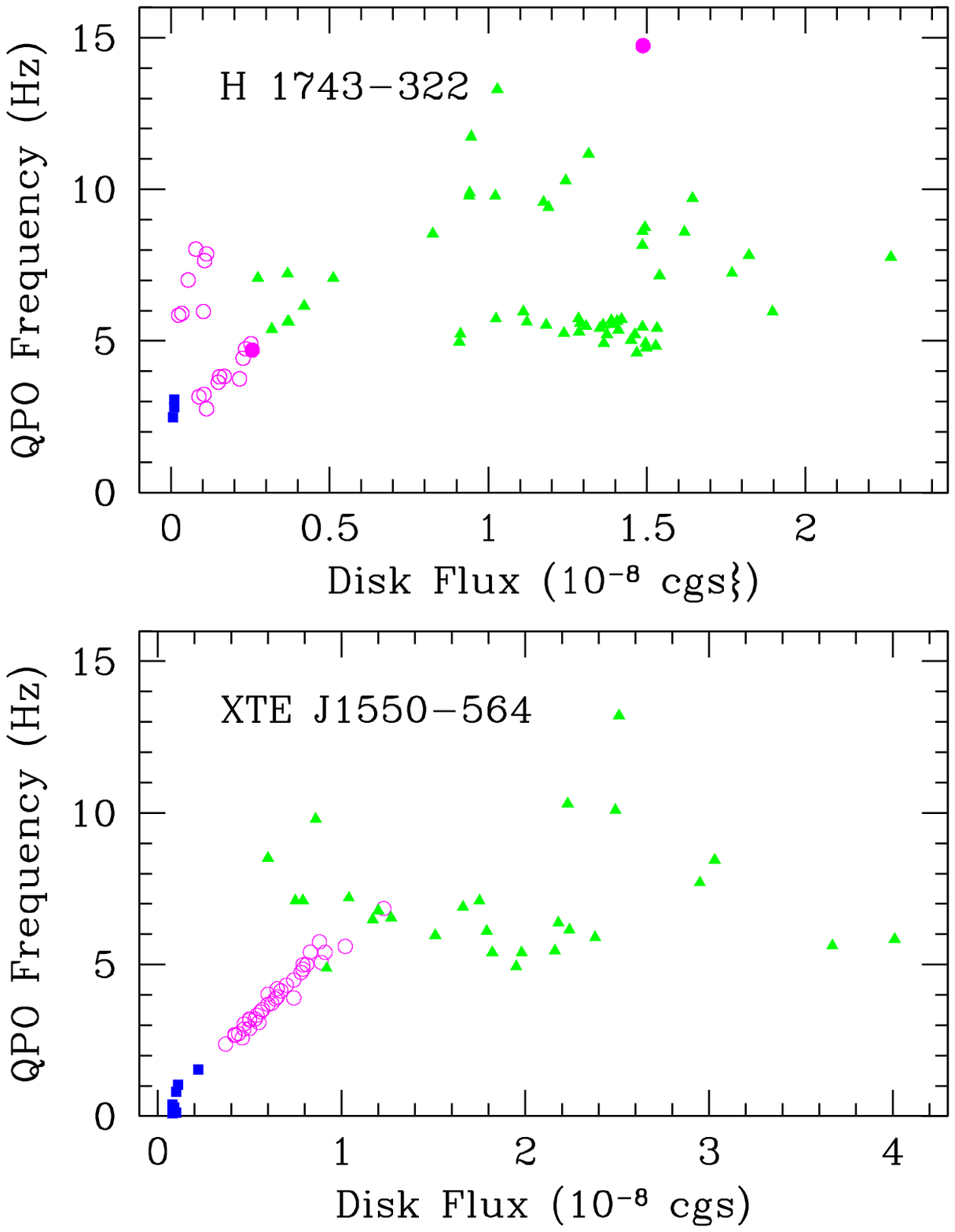] {QPO frequency versus disk flux for H1743 and XTE
  J1550--564.  For the key to the plotting symbols, see Figure 3.  For
  H1743, the strongly correlated data (see text) were obtained in the
  flaring phase during the first half of the outburst; the group of
  seven H:SPL data points near zero disk flux and between $\nu \sim
  6-8$~Hz are unrelated data obtained late in the outburst cycle (obs.\
  nos.\ 152, 154--159).  The error bars, which are smaller than or
  comparable to the size of the plotting symbols, are omitted for the
  sake of clarity.  One extreme data point for XTE J1550--564 for
  MJD51239.08 is omitted to maintain reasonable scales on the axes.
  (The flux errors for XTE J1550--564 were overestimated by Sobczak et
  al. 2000b by a factor of $\approx 5$.)}

\figcaption[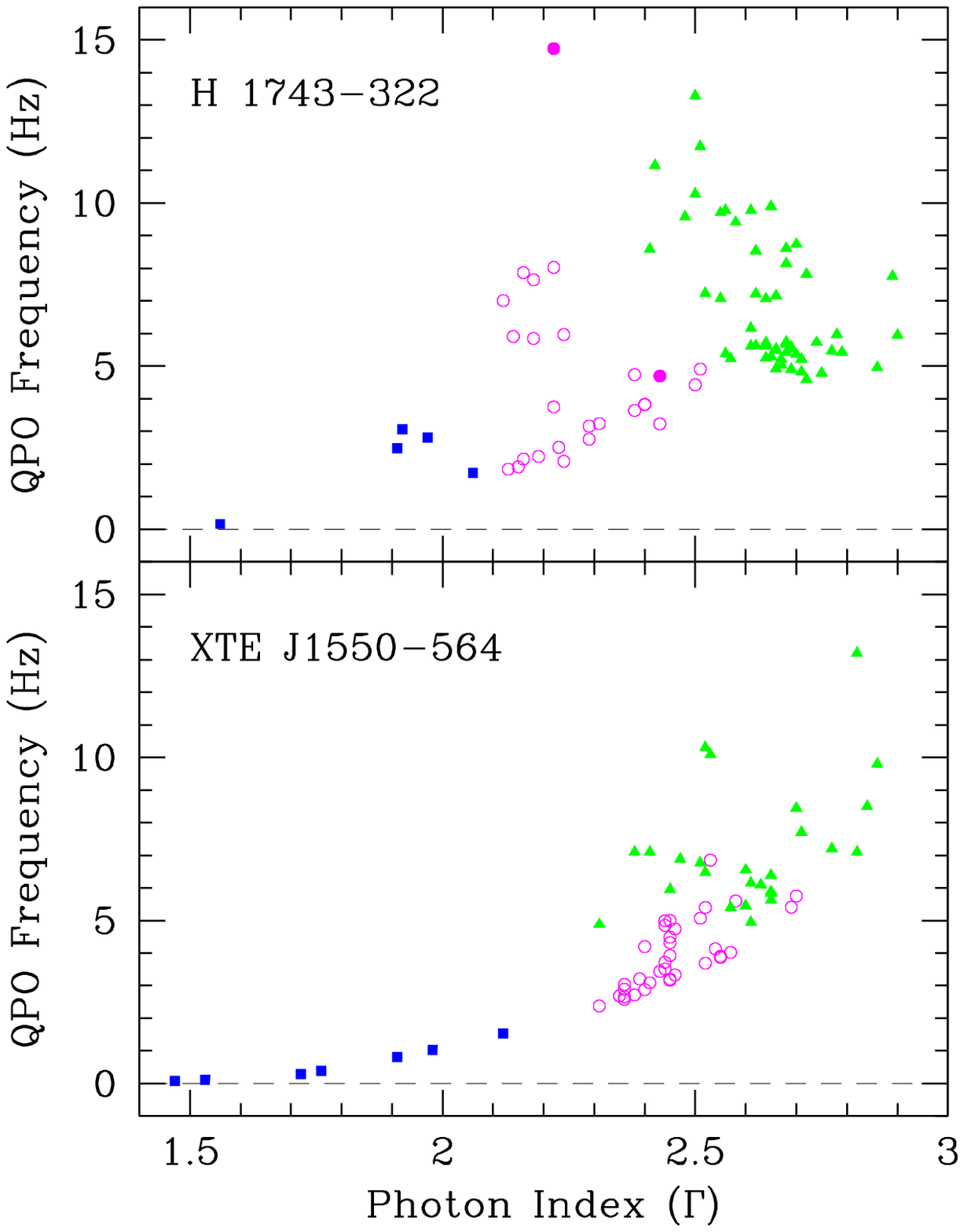] {QPO frequency versus the photon index $\Gamma$ ($E
  < E_{\rm break}$).  For the key to the plotting symbols, see Figure 3.}

\figcaption[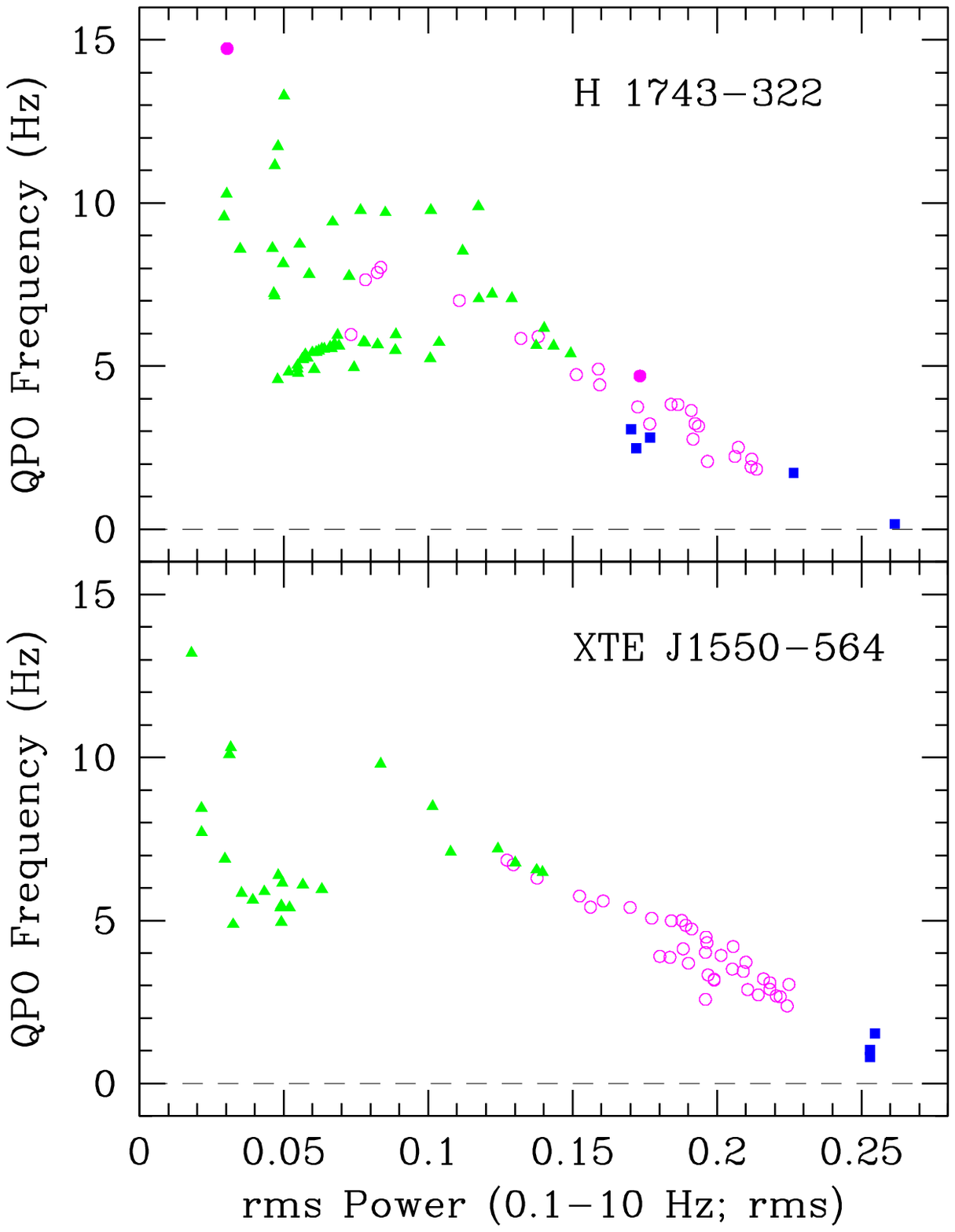] {QPO frequency versus the continuum power density
  (0.1--10 Hz).  For the key to the plotting symbols, see Figure 3.}

\figcaption[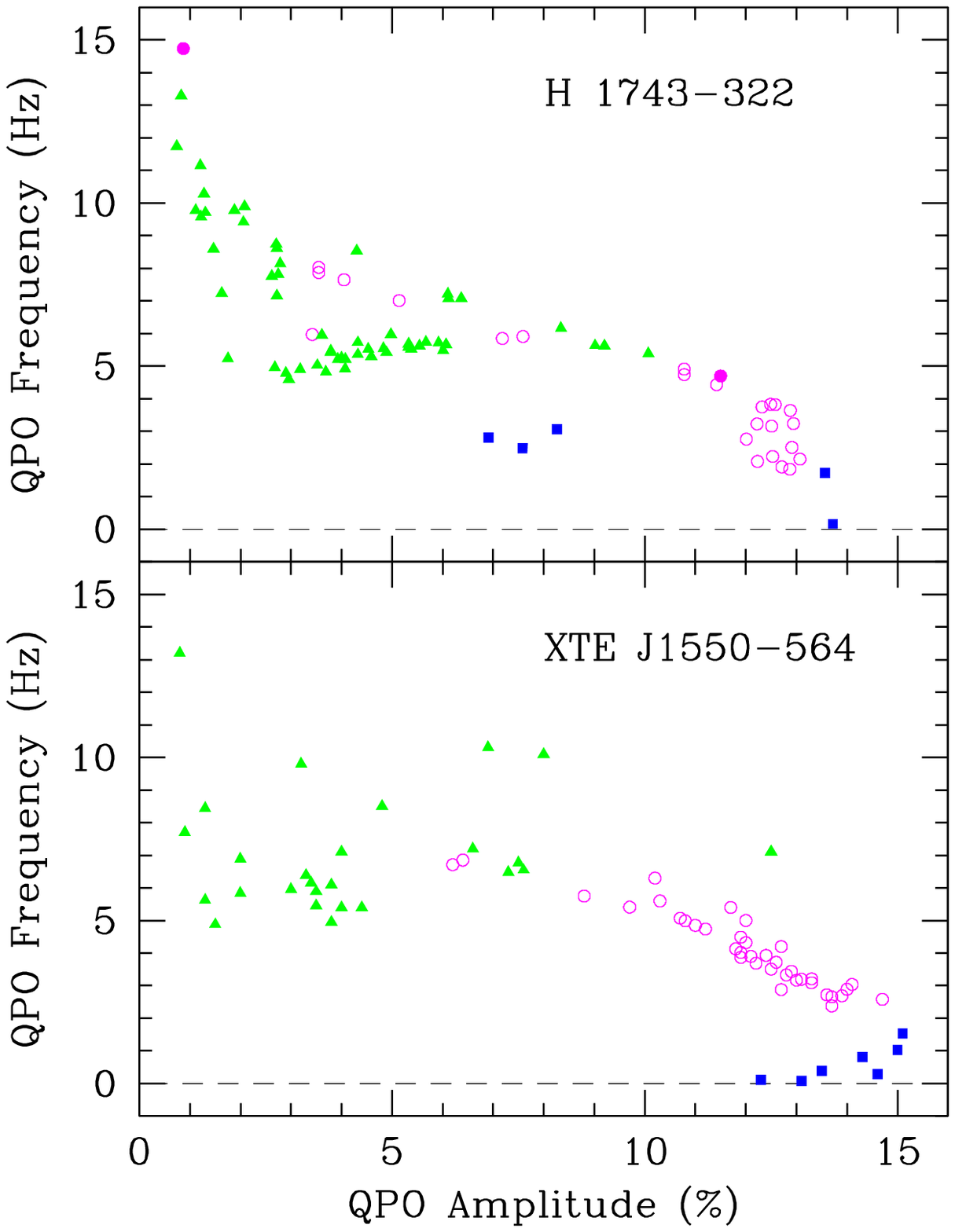] {QPO frequency versus QPO amplitude.  For the key
  to the plotting symbols, see Figure 3.}

\figcaption[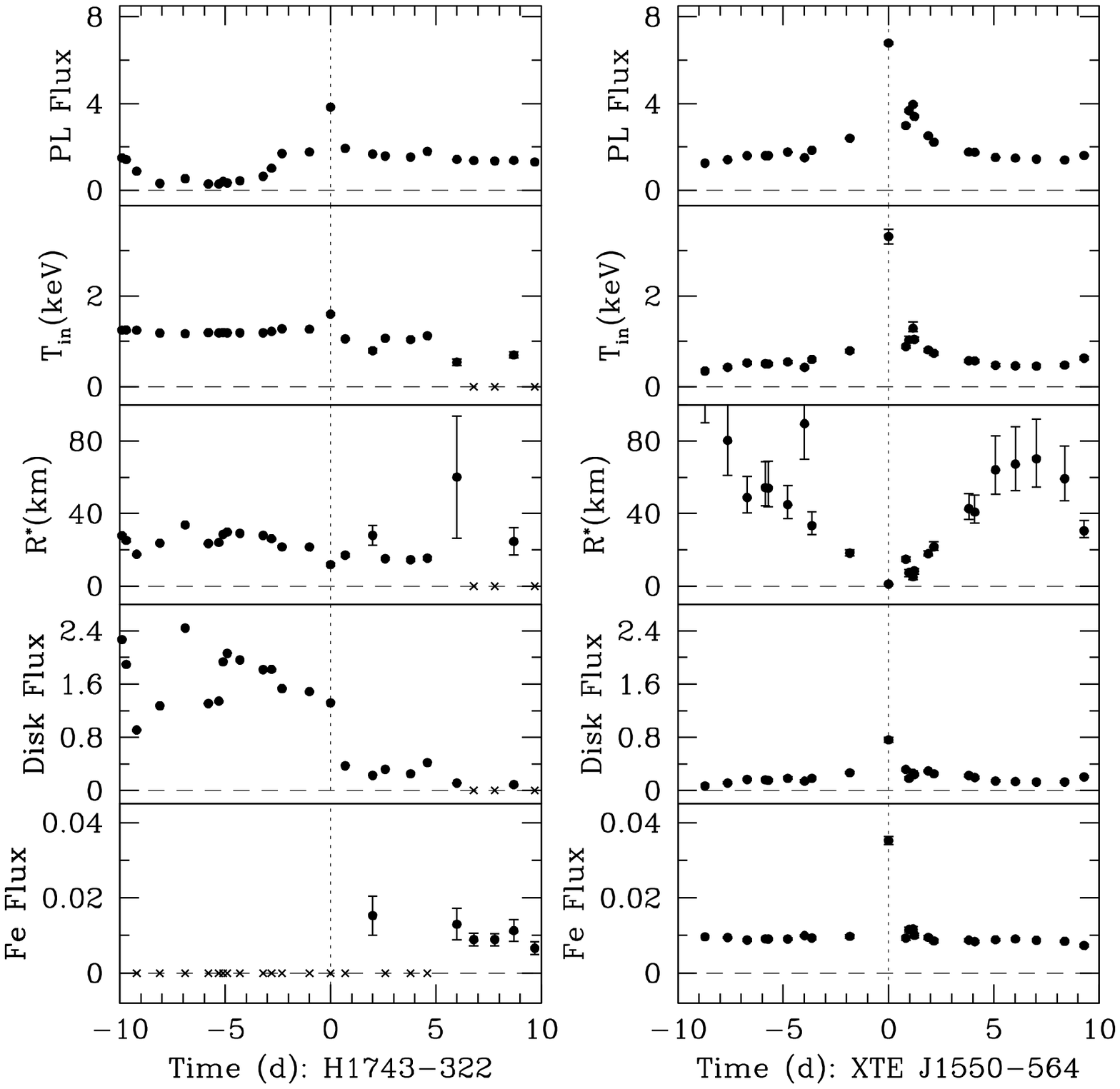] {Detailed look at the strongest power-law flares
  observed for H1743 (left panel) and XTE J1550--564 (right panel).
  From top to bottom are plotted the power-law flux, disk temperature,
  inner disk radius, disk flux, and Fe K line flux (the fluxes are in
  units of \flux) with the maximum of the power-law flux at time zero.
  To facilitate the comparison, the fluxes and radii for XTE J1550--564
  have been scaled down by the factors $(5.5/10)^2$ and (5.5/10),
  respectively, from the values given by Sobczak et al. (2000b) to
  correct for the lesser distance of this source (see \S3.3).}

\figcaption[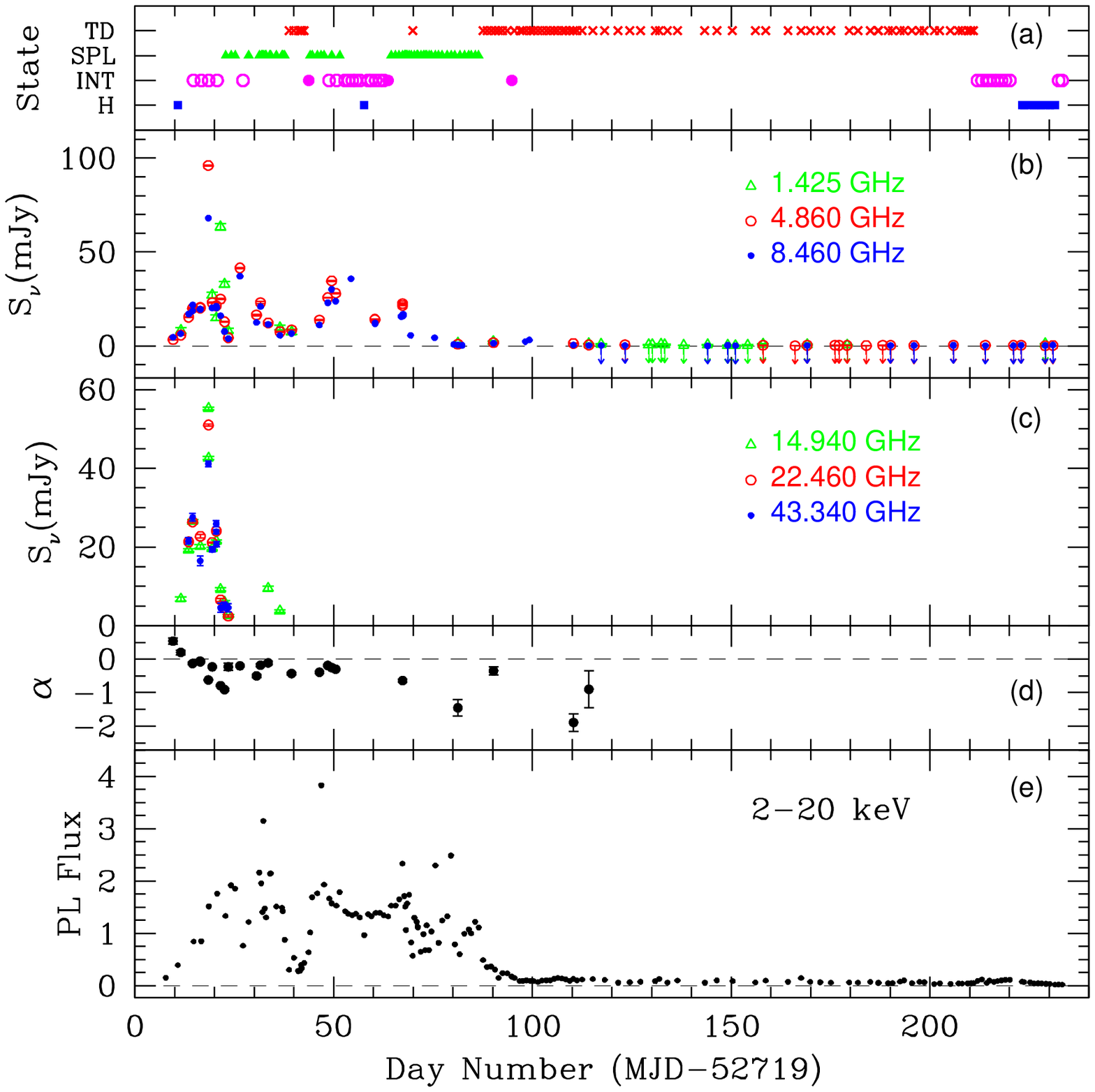] {VLA radio data.  The low-frequency data ($b$)
  extend over the entire outburst cycle (see Table A3), and the
  high-frequency data ($c$) are limited to the first 40 days (see Table
  A4).  The radio spectral index $\alpha$ ($S_{\rm \nu} \propto
  \nu^{\alpha}$), which is based on the 4.860 GHz and 8.460 GHz data, is
  shown in panel $d$, and the 2--20 keV power-law flux is shown in ($e$)
  for comparison.  The X-ray state is indicated in panel $a$ (see Fig.\
  3 for key to plotting symbols).}

\figcaption[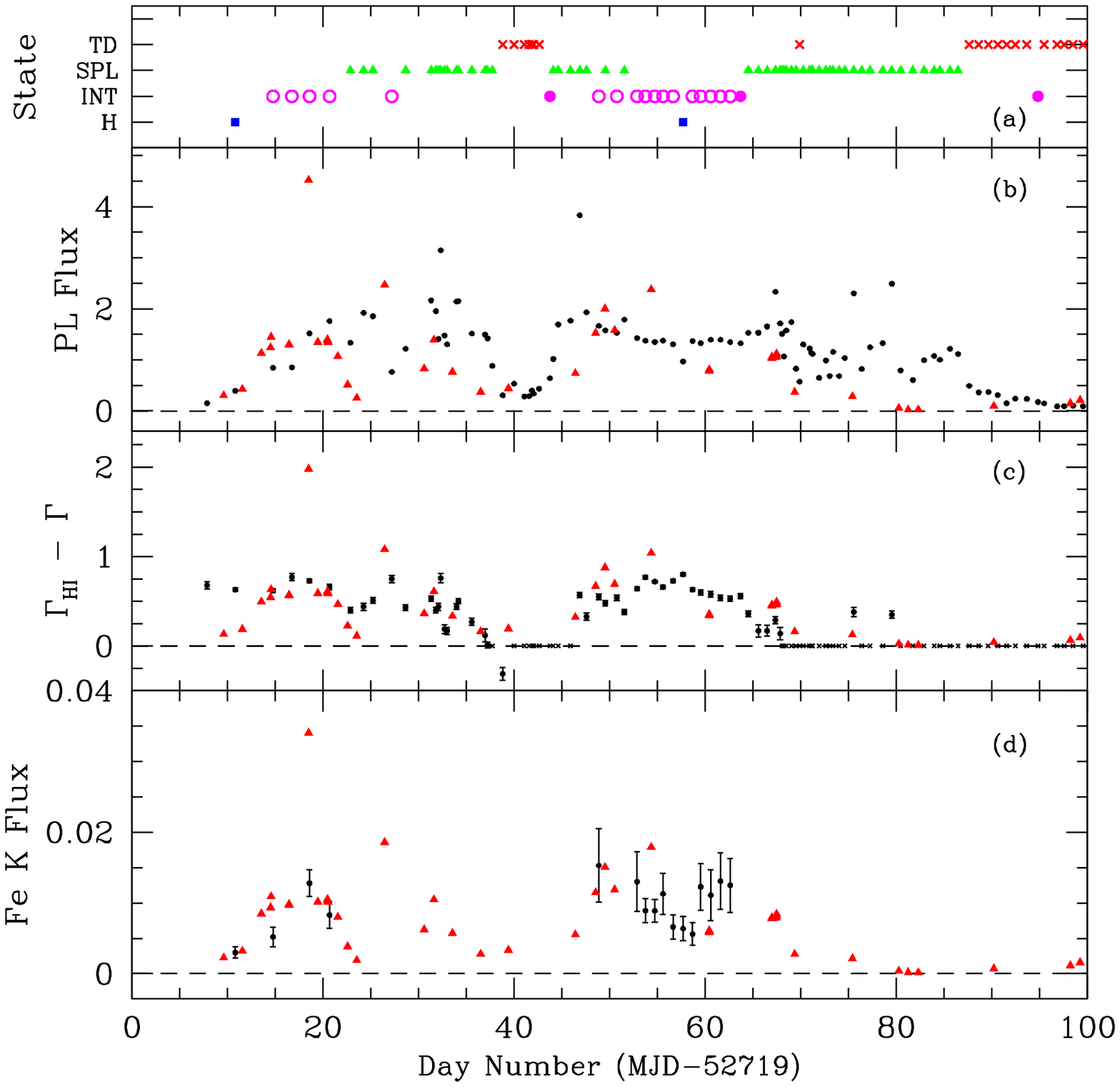] {Superposition of radio and X-ray data are shown
  for the first 100 days of the outburst cycle. The same 8.460 GHz radio
  data are arbitrarily scaled and plotted (red triangles) in each of the
  three panels. The X-ray data shown are ($b$) the 2--20 keV power-law
  flux, ($c$) the difference in the power-law indices $\Gamma$ and \GHI,
  and the Fe K line flux (see Figs.\ $4g$, $4d$ \& $4e$, respectively).
  The X-ray state is indicated in panel $a$ (see Fig.\ 3 for key to
  plotting symbols).}

\figcaption[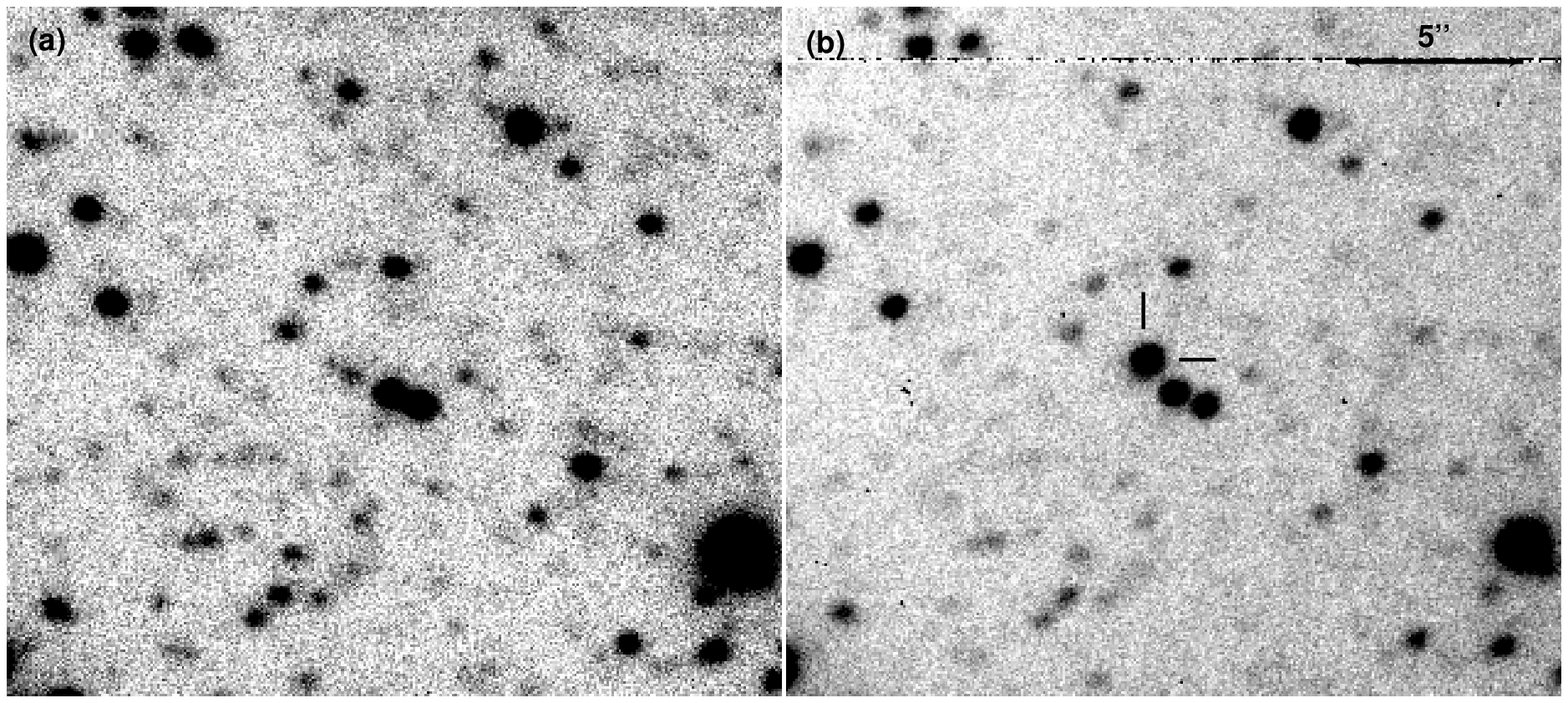] {($a$) Quiescent $i'$-band image obtained on
  2006 June 23 UT. The image is a median of five 300 s exposures
  obtained under 0\fas43 seeing.  ($b$) $I$-band image obtained near
  the start of the 2003 outburst on Day 15.4 (2003 April 5.4 UT). Tick
  marks point the location of the optical counterpart. The exposure
  time was 330 s and the seeing was 0\fas48.  North is up and East is
  left with a 20$''$ field of view.}

\figcaption[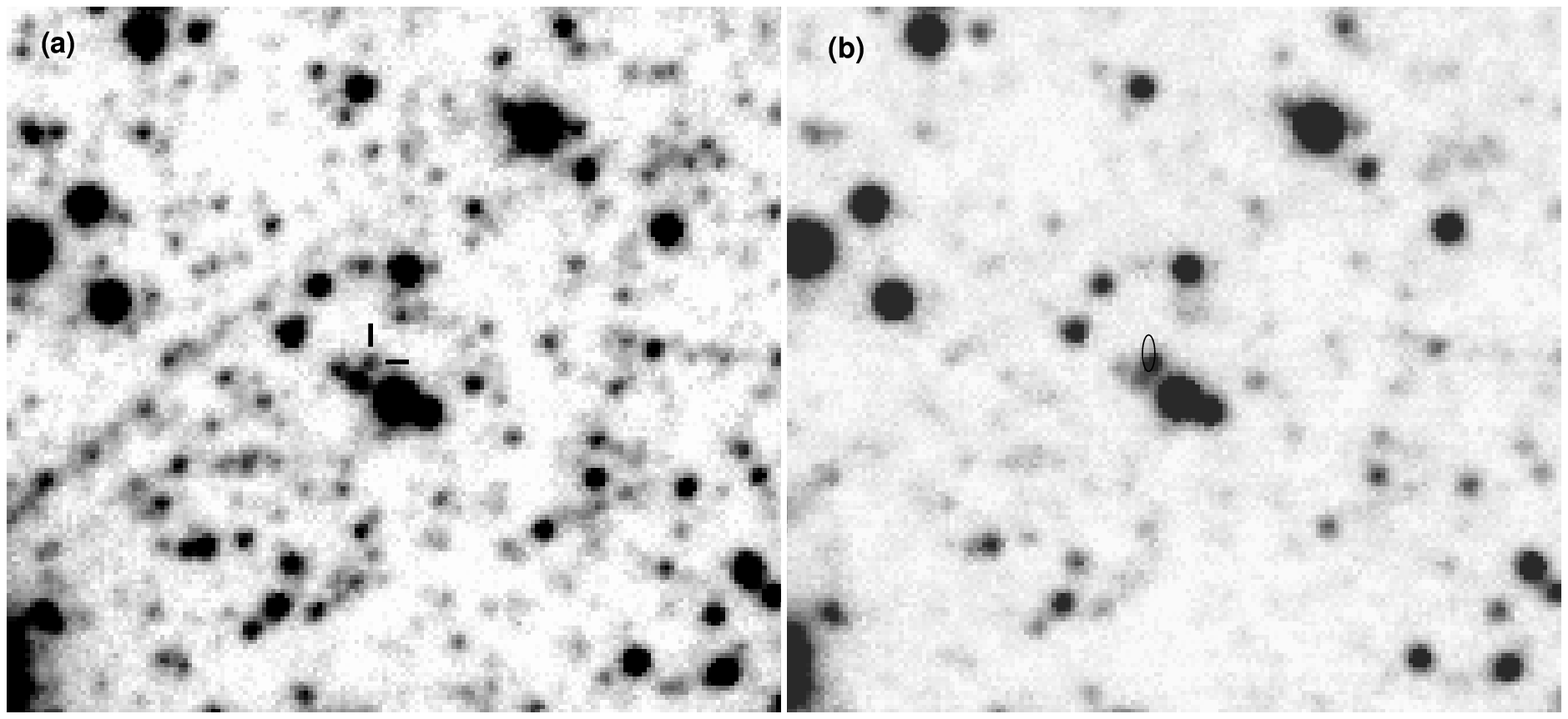] {$K$-band images.  ($a$) Quiescent image obtained
  on 2006 May 7.4 UT; the seeing is 0\fas43 and the exposure time is
  375~s.  Tick marks indicate the location of the near-infrared
  counterpart. ($b$) Image obtained near the end of the outburst on Day
  175.1 (2005 September 12.1 UT); the exposure time was 150 s and the
  seeing was 0\fas55. The error ellipse marks the position of the radio
  counterpart to H1743; the NIR counterpart is located near the bottom
  of the ellipse.  North is up and East is left.}

\clearpage
\begin{figure}
\figurenum{1} 
\plotone{f1.eps}
\caption{ }
\end{figure}

\clearpage
\begin{figure}
\figurenum{2}
\plotone{f2.eps}
\caption{ }
\end{figure}

\clearpage
\begin{figure}
\figurenum{3}
\plotone{f3.eps}
\caption{ }
\end{figure}

\clearpage
\begin{figure}
\figurenum{4}
\plotone{f4.eps}
\caption{ }
\end{figure}

\clearpage
\begin{figure}
\figurenum{5}
\plotone{f5.eps}
\caption{ }
\end{figure}

\clearpage
\begin{figure}
\figurenum{6}
\plotone{f6.eps}
\caption{ }
\end{figure}

\clearpage
\begin{figure}
\figurenum{7}
\plotone{f7.eps}
\caption{ }
\end{figure}

\clearpage
\begin{figure}
\figurenum{8}
\plotone{f8.eps}
\caption{ }
\end{figure}

\clearpage
\begin{figure}
\figurenum{9}
\plotone{f9.eps}
\caption{ }
\end{figure}

\clearpage
\begin{figure}
\figurenum{10}
\plotone{f10.eps}
\caption{ }
\end{figure}

\clearpage
\begin{figure}
\figurenum{11}
\plotone{f11.eps}
\caption{ }
\end{figure}

\clearpage
\begin{figure}
\figurenum{12}
\plotone{f12.eps}
\caption{ }
\end{figure}

\clearpage
\begin{figure}
\figurenum{13}
\plotone{f13.eps}
\caption{ }
\end{figure}

\clearpage
\begin{figure}
\figurenum{14}
\plotone{f14.eps}
\caption{ }
\end{figure}

\clearpage
\begin{figure}
\figurenum{15}
\plotone{f15.eps}
\caption{ }
\end{figure}

\clearpage
\begin{figure}
\figurenum{16}
\plotone{f16.eps}
\caption{ }
\end{figure}


\clearpage

\begin{thebibliography}{}

\bibitem[Abe et~al.(2005)]{abe05} Abe, Y., Fukazawa, Y., Kubota, A.,
Kasama, D., \& Makishima, K. 2005, PASJ, 57, 629

\bibitem[Abramowicz \& Kluzniak(2001)]{abr01}
Abramowicz, M.~A., \& Kluzniak, W. 2001, A\&A, 374, L19

\bibitem[Akaike(1974)]{aka74} Akaike, H. 1974, in Automatic Control,
IEEE Transactions, 19, 716

\bibitem[Arnaud(1996)]{arn96} Arnaud, K.~A. 1996, in A.\ S.\ P.\
Conference Series, Vol.\ 101, Astronomical Data Analysis Software and
Systems V, ed. G. H. Jacoby \& J. Barnes, 17

\bibitem[Baba et al.(2003)]{bab03}
Baba, D., \& Nagata, T. 2003, IAUC 8112

\bibitem[Balucinska-Church \& McCammon(1992)]{bal92}
Balucinski-Church, M., \& McCammon, D. 1992, ApJ, 400, 699

\bibitem[Belloni et al.(2005)]{bel05}
Belloni, T., Homan, J., Casella, P., van der Klis, M., Nespoli, E.,
Lewin, W.~H.~G., Miller, J.~M., \& M\'endez, M.  2005, A\&A, 440, 207

\bibitem[Brenneman \& Reynolds(2006)]{bre06}
Brenneman, L.~W., \& Reynolds, C.~S. 2006, ApJ, 652, 1028

\bibitem[Capitanio et al.(2005)]{cap05}
Capitanio, F., et al. 2005, ApJ, 622, 503

\bibitem[Capitanio et al.(2006)]{cap06}
Capitanio, F., et al. 2006, ApJ, 643, 376

\bibitem[Cooke et al.(1984)]{coo84}
Cooke, B.~A., Levine, A.~M., Lang, F.~L., Primini, F.~A., \& Lewin,
W.~H.~G. 1984, ApJ, 285, 258

\bibitem[Corbel et al.(2002)]{cor02} 
Corbel, S., Fender, R.~P., Tzioumis, A.~K., Tomsick, J.~A., Orosz,
J.~A., Miller, J.~M., Wijnands, R., \& Kaaret, P. 2002, Science, 298,
196

\bibitem[Corbel et al.(2005)]{cor05}
Corbel, S., Kaaret, P., Fender, R.~P., Tzioumis, A.~K., Tomsick, J.~A.,
\& Orosz, J.~A. 2005, ApJ, 632, 504

\bibitem[Davis et al.(2005)]{dav05}
Davis, S.~W., Blaes, O.~M., Hubeny, I., \& Turner, N.~J. 2005, ApJ, 621,
372

\bibitem[Davis et al.(2006)]{dav06}
Davis, S.~W., Done, C., \& Blaes, O.~M. 2006, ApJ, 647, 525

\bibitem[Doxsey et al.(1977)]{dox77}
Doxsey, R., et al. 1977, IAUC 3113

\bibitem[Ebisawa et al.(1994)]{ebi94}
Ebisawa, K., et al. 1994, PASJ, 46, 375

\bibitem[Fender et al.(2004)]{fen04}
Fender, R.~P., Belloni, T.~M., \& Gallo, E. 2004, MNRAS, 355, 1105

\bibitem[Gierli\'nski \& Done(2004)]{gie04}
Gierli\'nski, M. \& Done, C. 2004, MNRAS, 347, 885

\bibitem[Gou et al.(2009)]{gou09}
Gou, L., McClintock, J.~E., Liu, J., Narayan, R., Steiner, J.~F.,
Remillard, R.~A., Orosz, J.~A., \& Davis, S.~W. 2009, ApJ, submitted,
arXiv:0901.0920v1 (astro-ph)

\bibitem[Gursky et al.(1978)]{gur78} 
Gursky, H., et al. 1978, ApJ, 223, 973

\bibitem[Hannikainen et al.(2001)]{han01}
Hannikainen, D., Campbell-Wilson, D., Hunstead, R., McIntyre, V.,
Lovell, J., Reynolds, J., Tzioumis, T., \& Wu, K.  2001, Ap\&SSS, 276,
45

\bibitem[Homan et al.(2005)]{hom05}
Homan, J., Miller, J.~M., Wijnands, R., van der Klis, M., Belloni, T.,
Steeghs, D., \& Lewin, W.~H.~G. 2005, ApJ, 623, 383

\bibitem[Jahoda et al.(2006)]{jah06}
Jahoda, K., Markwardt, C.~B., Radeva, Y., Rots, A.~H., Stark, M.~J.,
Swank, J.~H., Strohmayer, T.~E., \& Zhang, W. 2006, ApJS, 163, 401

\bibitem[Joinet et al.(2005)]{joi05}
Joinet, A., Jourdain, E., Malzac, J., Roques, J.~P., Sch\"onfelder, V.,
Ubertini, P., \& Capitanio, F. 2005, ApJ, 629, 1008

\bibitem[Kaaret et al.(2003)]{kaa03}
Kaaret, P., Corbel, S., Tomsick, J.~A., Fender, R., Miller, J.~M.,
Orosz, J.~A., Tzioumis, A.~K., \& Wijnands, R. 2003, ApJ, 582, 945

\bibitem[Kalemci et al.(2006)]{kal06}
Kalemci, E., Tomsick, J.~A., Rothschild, R.~E., Pottschmidt, K., Corbel,
S., \& Kaaret, P. 2006, ApJ, 639, 340

\bibitem[Kaluzienski \& Holt(1977)]{kal77}
Kaluzienski, L.~J., \& Holt, S.~S. 1977, IAUC 3099

\bibitem[Kubota \& Makishima(2004)]{kum04}
Kubota, A., \& Makishima, K. 2004, ApJ, 601, 428

\bibitem[Kubota et al.(2001)]{kub01}
Kubota, A., Makishima, K., \& Ebisawa, K. 2001, ApJ, 560, L147

\bibitem[Liu et~al.(2008)]{liu08}
Liu, J., McClintock, J.~E., Narayan, R., Davis, S.~W., \& Orosz,
J.~A. 2008, ApJ, 679, L37

\bibitem[Lutovinov et al.(2005)]{lut05} 
Lutovinov, A., Revnivtsev, M., Molkov, S., \& Sunyaev, R. 2005, A\&A,
430, 997

\bibitem[Makishima et al.(1986)]{mak86}
Makishima, K., Maejima, Y., Mitsuda, K., Bradt, H.~V., Remillard, R.~A.,
Tuohy, I.~R., Hoshi, R., \& Nakagawa, M. 1986, ApJ, 308, 635

\bibitem[Markwardt \& Swank(2003)]{mar03}
Markwardt, C.~B., \& Swank, J.~H. 2003, Astron.\ Telegram 133, 136

\bibitem[Markwardt et al.(1999)]{mar99}
Markwardt, C.~B., \& Swank, J.~H., \& Taam, R.~E. 1999, ApJ, 513, L37

\bibitem[Martini et al.(2004)]{mar04} 
Martini, P., Persson, S.~E., Murphy, D.~C., Birk, C., Shectman, S.~A.,
Gunnels, S.~M., \& Koch, E. 2004, Proc. SPIE, 5492, 1653

\bibitem[McClintock \& Remillard(2006)]{mcc06a}
McClintock, J.~E., \& Remillard, R.~A. 2006, in Compact
Stellar X-ray Sources, eds. W. Lewin \& M. van der Klis (Cambridge:
Cambridge Univ. Press), p157 

\bibitem[McClintock et al.(2006)]{mcc06b}
McClintock, J.~E., Shafee, R., Narayan, R., Remillard, R.~A., Davis,
S.~W., \& Li, L.-X. 2006, ApJ, 652, 518

\bibitem[McClintock et al.(2007)]{mcc07}
McClintock, J.~E., Narayan, R., \& Shafee, R. 2007, to appear in Black
Holes, ed.\ M. Livio and A. Koekemoer, (Cambridge University Press:
Cambridge), arXiv:0707.4492v1 (astro-ph)

\bibitem[Miller et al.(2006)]{mil04}
Miller, J.~M., et al. 2006, ApJ, 646, 394

\bibitem[Miller et al.(2008)]{mil08} 
Miller, J.~M., et al. 2008, ApJ, 679, 113

\bibitem[Mitsuda et al.(1984)]{mit84}
Mitsuda, K., et al. 1984, PASJ, 36, 741

\bibitem[Muno et al.(1999)]{mun99}
Muno, M.~P., Morgan, E.~H., \& Remillard, R.~A. 1999, ApJ, 527, 321

\bibitem[Narayan \& McClintock(2005)]{nar05}
Narayan, R., \& McClintock, J.~E. 2005, ApJ, 623, 1017

\bibitem[Orosz et al.(2002)]{oro02}
Orosz, J. A., et al. 2002, ApJ, 568, 2002

\bibitem[Orosz et~al.(2007)]{oro07}
Orosz, J.~A., et al. 2007, Nature, 449, 872

\bibitem[Park et al.(2004)]{par04}
Park, S.~Q., et al. 2004, ApJ, 610, 378

\bibitem[Parmar et al.(2003)]{par03}
Parmar, A.~N., Kuulkers, E., Oosterbroek, P., Barr, P., Much, R., Orr,
A., Williams, O.~R., \& Winkler, C.  2003, A\&A, 411, L421

\bibitem[Persson et al.(1998)]{per98}
Persson, S.~E., Murphy, D.~C., Krzeminski, W., Roth, M., \& Rieke, M.~J.
1998, AJ, 116, 2475

\bibitem[Predehl \& Schmitt(1995)]{pre95}
Predehl, P., \& Schmitt, J.~H.~M.~M. 1995, A\&A, 293, 889

\bibitem[Reis et al.(2008)]{rei08} Reis, R.~C., Fabian, A.~C., Ross,
R.~R., Miniutti, G., Miller, J.~M., \& Reynolds, C. 2008, MNRAS, 387,
1489

\bibitem[Remillard \& McClintock(2006)]{rem06}
Remillard, R.~A., \& McClintock, J.~E. 2006, ARAA, 44, 49 (RM06)

\bibitem[Remillard et al.(2006)]{rem04}
Remillard, R.~A., McClintock, J.~E., Orosz, J.~A., \& Levine A.~M. 2006,
ApJ, 637, 1002

\bibitem[Remillard et al.(2002a)]{rem02a}
Remillard, R.~A., Muno, M.~P., McClintock, J.~E., \& Orosz, J.~A. 2002a,
ApJ, 580, 1030

\bibitem[Remillard et al.(2002b)]{rem02b}
Remillard, R.~A., Sobczak, G.~J., Muno, M.~P., \& McClintock,
J.~E. 2002b, ApJ, 564, 962

\bibitem[Revnivtsev et al.(2003)]{rev03}
Revnivtsev, M., Chernyakova, M., Westergaard, N.~J., Shoenfelder, V.,
Gehrels, N., \& Winkler, C. 2003, Astron.\ Telegram, 132

\bibitem[Revnivtsev et al.(2000)]{rev05}
Revnivtsev, M.~G., Trudolyubov, S.~P., \& Borozdin, K.~N. 2000, MNRAS,
312, 151

\bibitem[Rothschild et al.(1998)]{rot98}
Rothschild, R.~E., et al. 1998, ApJ, 496, 538

\bibitem[Rupen et al.(2003)]{rup03}
Rupen, M.~P., Mioduszewski, A.~J., \& Dhawan, V. 2003, Astron.\
Telegram, 137, 139

\bibitem[Schlegel et al.(1998)]{sch98} 
Schlegel, D.~J., Finkbeiner, D.~P., \& Davis, M.  1998, ApJ, 500, 525

\bibitem[Shafee et~al.(2006)]{sha06}
Shafee, R., McClintock, J.~E., Narayan, R., Davis, S.~W., Li, L.-X.,
\& Remillard, R.~A. 2006, ApJL, 636, L113

\bibitem[Shakura \& Sunyaev(1973)]{sha73}
Shakura, N.~I., \& Sunyaev, R.~A. 1973, A\&A, 24, 337

\bibitem[Shimura \& Takahara(1995)]{shi95}
Shimura, T., \& Takahara, F. 1995, ApJ, 445, 780

\bibitem[Silverman \& Filippenko(2008)]{sil08}
Silverman, J.~M., \& Filippenko A.~V. 2008, ApJ, 678, L17

\bibitem[Skrutskie et~al.(2006)]{skr06}
Skrutskie et al.\ 2006, AJ, 131, 1163

\bibitem[Sobczak et al.(1999)]{sob99}
Sobczak, G.~J., McClintock, J.~E., Remillard, R.~A., \& Bailyn,
C.~D. 1999, ApJ, 520, 776

\bibitem[Sobczak et al.(2000a)]{sob00a}
Sobczak, G.~J., McClintock, J.~E., Remillard, R.~A., Cui, W., Levine,
A.~M., \& Morgan, E.~H. 2000a, ApJ, 531, 537

\bibitem[Sobczak et al.(2000b)]{sob00b}
Sobczak, G.~J., McClintock, J.~E., Remillard, R.~A., Cui, W., Levine,
A.~M., Morgan, E.~H., Orosz, J.~A., \& Bailyn, C.~D.  2000b, ApJ, 544,
993

\bibitem[Steeghs et al.(2003ab)]{ste03} 
Steeghs, D., Miller, J.~M., Kaplan, D., \& Rupen, M. 2003, Astron.\
Telegram 141, 146

\bibitem[Stetson(1987)]{ste87}
Stetson, P.~B. 1987, PASP, 99, 191

\bibitem[Tanaka \& Lewin(1995)]{tan95}
Tanaka, Y., \& Lewin, W.~H.~G. 1995, in X-ray Binaries, eds. W. Lewin, 
J. van Paradijs, \& E. van den Heuvel (Cambridge: Cambridge Univ. Press),
p126

\bibitem[Titarchuk et al.(2004)]{tit04}
Titarchuk, L., \& Fiorito, R. 2004, ApJ, 612, 988

\bibitem[Titarchuk et al.(2005)]{tit05}
Titarchuk, L., \& Shaposhnikov, N. 2005, ApJ, 626, 298

\bibitem[Tomsick et al.(2003)]{tom03}
Tomsick, J.~A., Corbel, S., Fender, R., Miller, J.~M., Orosz, J.~A.,
Tzioumis, T., Wijnands, R., \& Kaaret, P. 2003, ApJ, 582, 933 

\bibitem[Tomsick et al.(2005)]{tom05}
Tomsick, J.~A., Corbel, S., Goldwurm, A., \& Kaaret, P. 2005, ApJ, 
630, 413

\bibitem[Van Paradijs \& McClintock(1994)]{par94}
van Paradijs, J., \& McClintock, J.~E. 1994, A\&A, 290, 133

\bibitem[Vignarca et al.(2003)]{vig03}
Vignarca, F., Migliari, S., Belloni, T., Psaltis, D., \& van der Klis,
M. 2003, A\&A, 397, 729

\bibitem[Wagoner(1999)]{wag99}
Wagoner, R.~V. 1999, Physics Reports, 311, 259

\bibitem[White \& Marshall(1984)]{whi84}
White, N.~E., \& Marshall, F.~E. 1984, ApJ, 281, 354

\bibitem[Wilms et al.(2006)]{wil06}
Wilms, J., Nowak, M.~A., Pottschmidt, K., Pooley, G.~G., \& Fritz, S.
2006, A\&A, 447, 245

\bibitem[Yan et al.(1998)]{yan98}
Yan, M., Sadeghpour, H.~R., \& Dalgarno, A. 1998, ApJ, 496, 1044

\end{thebibliography}
\end{document}